\documentclass[aps,prl,reprint,superscriptaddress,longbibliography,nobalancelastpage,groupedaddress]{revtex4-2}
\usepackage{graphicx}
\usepackage{dcolumn}
\usepackage{amsmath,amssymb}
\usepackage{latexsym}
\usepackage{mathrsfs}
\usepackage[usenames,dvipsnames]{color}
\usepackage{hyperref}
\usepackage{xcolor}

\usepackage[normalem]{ulem}
\newcommand{\be}{\begin{equation}} 
\newcommand{\ee}{\end{equation}} 
\newcommand{\bea}{\begin{eqnarray}} 
\newcommand{\eea}{\end{eqnarray}} 

\newcommand{\figref}[2]{[Fig.~\hyperref[#1]{\ref*{#1}(#2)}]}
\newcommand{\figrefi}[2]{[Fig.~\hyperref[#1]{\ref*{#1}(#2)}, inset]}
\newcommand{\textfigref}[2]{Fig.~\hyperref[#1]{\ref*{#1}(#2)}}
\newcommand{\textfigureref}[2]{Figure~\hyperref[#1]{\ref*{#1}(#2)}}

\newcommand{\figrefp}[2]{\hyperref[#1]{\ref*{#1}(#2)}}

\definecolor{linkcolor}{HTML}{223096}
\hypersetup{colorlinks,allcolors=linkcolor}

\bibpunct{\textcolor{linkcolor}{[}}{\textcolor{linkcolor}{]}}{\textcolor{linkcolor}{,}}{n}{}{;}
\renewcommand{\eqref}[1]{\hyperref[#1]{(\ref*{#1})}}
\newcommand{\Lambdai}{\mathit{\Lambda}}

\begin{document}
\title{(Un)buckling mechanics of epithelial monolayers under compression}

\author{Chandraniva \surname{Guha Ray}}
\affiliation{Max Planck Institute for the Physics of Complex Systems, N\"othnitzer Stra\ss e 38, 01187 Dresden, Germany}
\affiliation{\smash{Max Planck Institute of Molecular Cell Biology and Genetics, Pfotenhauerstra\ss e 108, 01307 Dresden, Germany}}
\affiliation{Center for Systems Biology Dresden, Pfotenhauerstra\ss e 108, 01307 Dresden, Germany}
\author{Pierre A. Haas}
\email{haas@pks.mpg.de}
\affiliation{Max Planck Institute for the Physics of Complex Systems, N\"othnitzer Stra\ss e 38, 01187 Dresden, Germany}
\affiliation{\smash{Max Planck Institute of Molecular Cell Biology and Genetics, Pfotenhauerstra\ss e 108, 01307 Dresden, Germany}}
\affiliation{Center for Systems Biology Dresden, Pfotenhauerstra\ss e 108, 01307 Dresden, Germany}
\date{\today}
\begin{abstract}
When cell sheets fold during development, their apical or basal surfaces constrict and cell shapes approach the geometric singularity in which these surfaces vanish. Here, we reveal the mechanical consequences of this geometric singularity for tissue folding in a minimal vertex model of an epithelial monolayer. In simulations of the buckling of the epithelium under compression and numerical solutions of the corresponding continuum model, we discover an ``unbuckling'' bifurcation: At large compression, the buckling amplitude can decrease with increasing compression. By asymptotic solution of the continuum equations, we reveal that this bifurcation comes with a large stiffening of the epithelium. Our results thus provide the mechanical basis for absorption of compressive stresses by tissue folds such as the cephalic furrow during germband extension in \emph{Drosophila}.
\end{abstract}\maketitle
During development, tissues fold into ``wrinkly'' shapes. Examples include the cortical convolutions in the brain \cite{Richman75,nie10,bayly13,tallinen14,Manyuhina14,Silvia15,tallinen16,lejeune16,holland20}, the intestinal villi \cite{hannezo11,Savin11,shyer13}, and the airways of the lungs \cite{Metzger08,Herriges14,Kim15,varner15}. This folding is crucial for tissue and organ function: The human brain, for example, is much more wrinkled than that of other primates \cite{Rogers10,Zilles13}, and lack of wrinkling in lissencephalies leads to severe developmental delays \cite{Walker42}. These tissue folding processes can be understood in terms of mechanical instabilities of wrinkling and buckling~\cite{nelson16}: For example, during brain development, the growth of the cortex of grey \mbox{matter~\cite{chenn02,neal07,xu10,stahl13,heide20}} exceeds that of the underlying white matter~\cite{Richman75,tallinen14,toro05}, which is thought to cause a geometric incompatibility that drives a wrinkling \mbox{instability~\cite{Manyuhina14,tallinen14,Silvia15,tallinen16,lejeune16,mahadevan13,engstrom18,holland18}} from which the gyrations of the brain emerge.

\begin{figure}[b]
\centering 
\includegraphics[width=8.6cm]{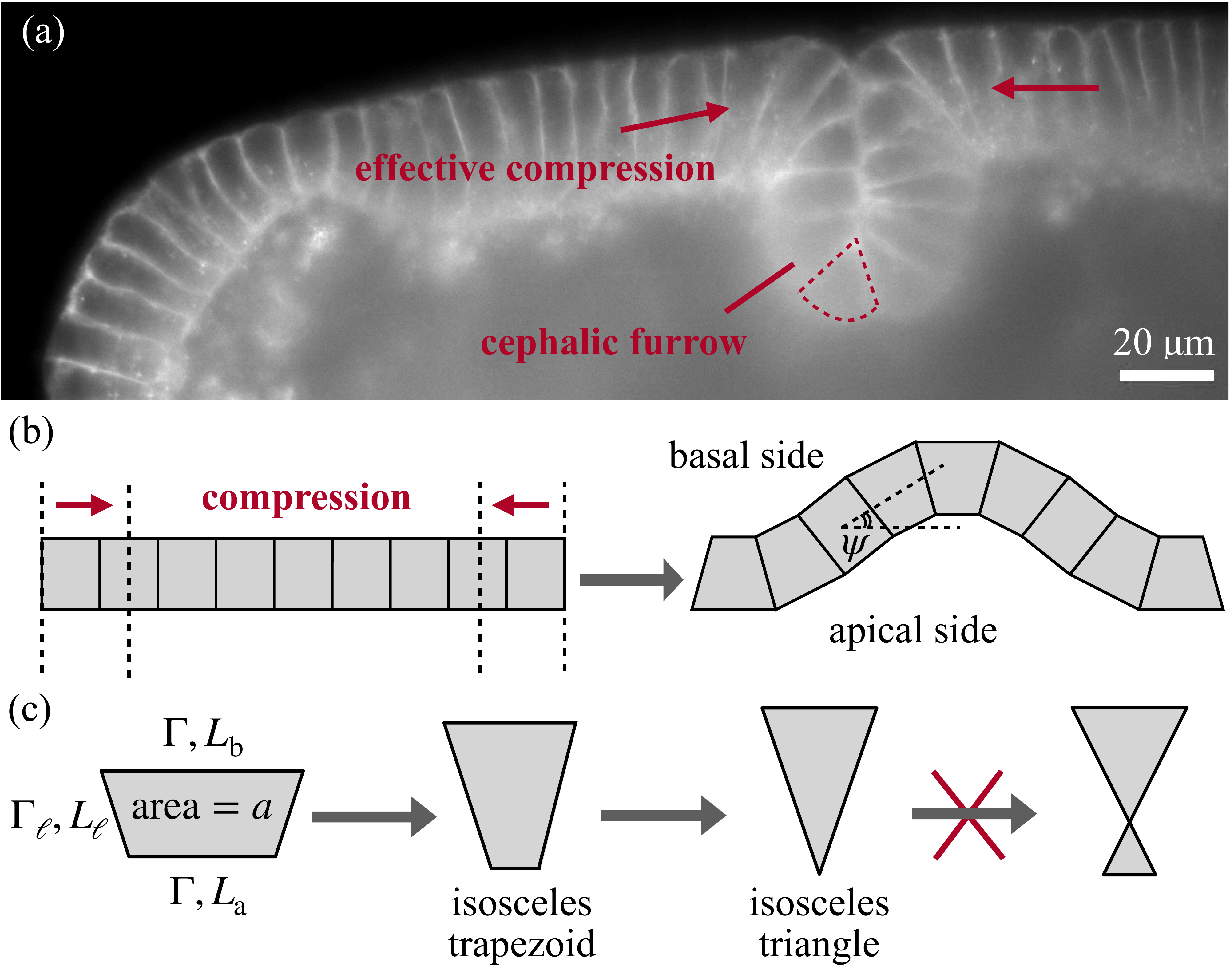}
\caption{Mechanics of tissue folds. (a)~Cephalic furrow in \emph{Drosophila}~\cite{vellutini23} with constricted triangular cell shapes. Arrows show the effective tissue compression from germband extension after the active formation of the furrow. Dotted lines outline one constricted cell. Image provided by Bruno Vellutini. Scale bar: 20 \textmu m. (b)~Cartoon of buckling of a tissue under external compression: As the tissue folds, the initially rectangular cells become trapezoidal. The apical and basal sides of the tissue, and the tangent angle $\psi$ of the tissue midline are also shown. (c) Geometric singularity of constriction in a minimal vertex model: Cell shapes are isosceles trapezoids of respective apical, basal, and lateral sides $L_\text{a},L_\text{b},L_\ell$ and fixed area $a$, with equal apical and basal tensions $\Gamma$ and lateral tension $\Gamma_\ell$. As the sides bend, the cells become triangular, at which point their lateral sides cannot bend further.}
\label{fig1}
\end{figure}

A very recent example~\cite{vellutini23,dey23} of the function of tissue folds concerns the cephalic furrow, an epithelial fold that forms actively during \emph{Drosophila} gastrulation~\figref{fig1}{a}. In mutants in which formation of the cephalic furrow is prevented, ectopic folds appear via a buckling instability caused by the concurrent tissue movements of germband extension~\cite{vellutini23}, suggesting that the cephalic furrow in the wild type absorbs the associated compressive stresses. There is however one key difference between this buckling of epithelial monolayers and the classical buckling of elastic rods or sheets \cite{landaulifshitz,audoly}: As a tissue folds, the lateral cell sides bend~\figref{fig1}{b}. Consequently, the apical or basal cell sides shrink until cells become triangular~\figref{fig1}{c}, at which point the cell sides cannot bend further. This constriction is a classical driver of tissue folding~\cite{sawyer10,keller11} that also appears in the cephalic furrow~\figref{fig1}{a}, but triangular cells constitute a geometric singularity that is absent from the classical buckling instabilities.

Here, we analyse the mechanical consequences of this geometric singularity in a minimal vertex model of a compressed epithelial monolayer. In simulations of this discrete model and in numerical solutions of its continuum limit, we reveal an ``unbuckling'' bifurcation: As the compression increases, the amplitude of the buckled monolayer can decrease. We analyse this bifurcation by solving the continuum equations asymptotically to reveal a large stiffening of the epithelium associated with the geometric singularity of cell constriction.

\begin{figure*}
\centering 
\includegraphics[width=17.8cm]{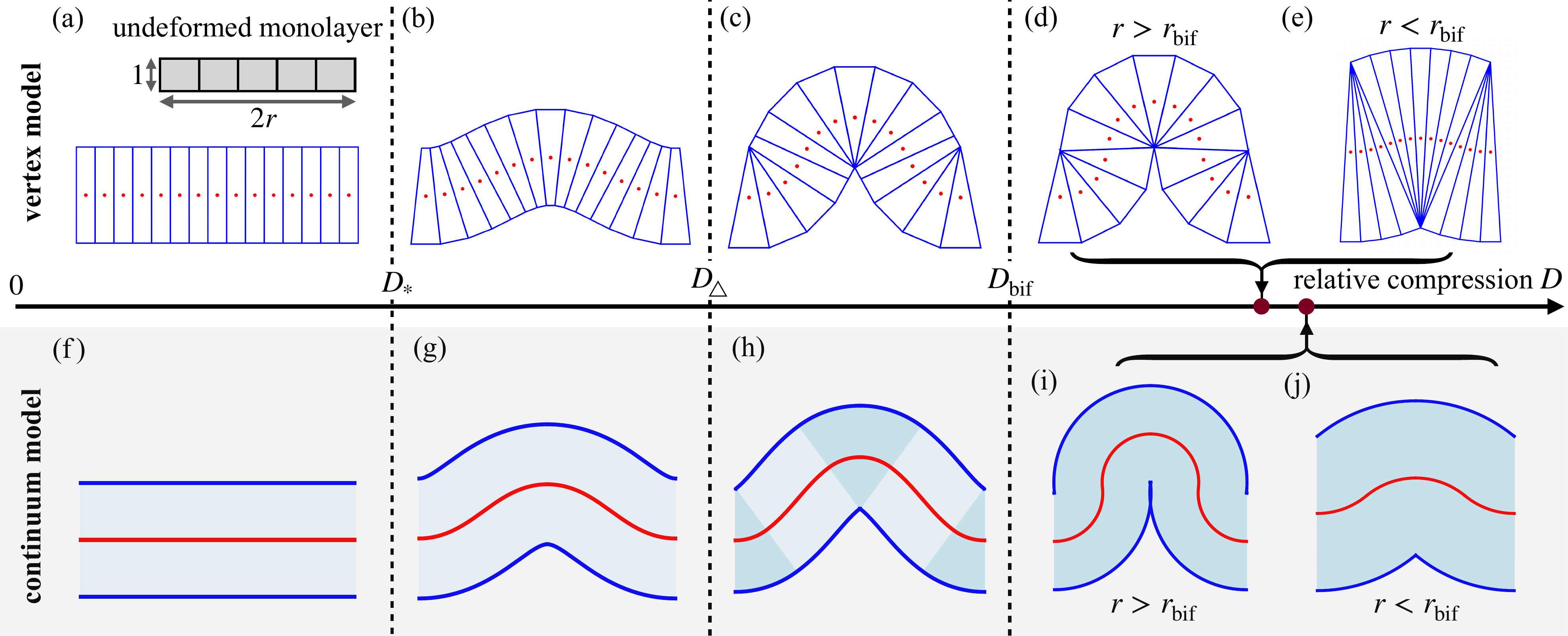}
\caption{(Un)buckling of a compressed epithelial monolayer as its relative compression $D$ is increased, in a minimal vertex model (top row) and in the corresponding continuum limit (bottom row). (a)~Flat configuration of the discrete monolayer before buckling. Red dots mark cell centres. Inset: definition of the aspect ratio $r$ from the undeformed shape of the monolayer. (b)~At $D=D_\ast$, the monolayer buckles. As $D>D_\ast$ increases, so does the buckling amplitude, and the cell sides bend. (c)~At $D=D_\triangle$, cells in the crest and the trough of the buckled shape become triangular, and, as $D>D_\triangle$ increases, fans of triangular cells (darker shading) expand from the crest and the trough. For still larger compressions $D>D_\text{bif}$, all cells are triangular and unbuckling is possible depending on the aspect ratio of the monolayer: (d)~If $r>r_\text{bif}$, the buckling amplitude continues to increase with $D$. (e)~If $r<r_\text{bif}$ however, at the same compression, the amplitude decreases with increasing $D$, and the triangular cells become taller and thinner. (f)--(j) Corresponding shapes in the continuum limit. The red line marks the midline of the cells, and the blue lines their apical and basal surfaces. Kinks in the apical or basal surfaces for $D>D_\triangle$ herald a fan of triangular cells (darker shading) at their crests and troughs.}
\label{fig2}
\end{figure*}

Our starting point is a vertex model of the two-dimensional cross-sections of epithelial cells \cite{derganc09, matej12, matej13, hannezo14, matej15, haas19, matej20, matej21, matej23}: The cells are isosceles trapezoids of fixed area $a=1$~\figref{fig1}{c}, with apical, basal, and lateral side lengths $L_{\text{a}}$, $L_\text{b}$, and $L_\ell$, respectively. The energy of an individual cell is
\begin{align}
e=\Gamma\bigl(L_\text{a}+L_\text{b}\bigr)+\Gamma_\ell L_\ell,\label{eq:sce}
\end{align}
where $\Gamma$ and $\Gamma_\ell$ are its apicobasal and lateral tensions~\footnote{Less minimal versions of these ``differential-tension'' models assign different tensions $\Gamma_\text{a}\neq\Gamma_\text{b}$ to the apical and basal cell sides~\cite{derganc09,matej12,matej13,hannezo14}. At sufficiently large $|\Gamma_\text{a}-\Gamma_\text{b}|$, these models undergo a wrinkling instability even in the absence of external compression~\cite{matej13,matej23}.}. We subject a flat monolayer of such cells \figref{fig2}{a} to external, relative compression $D$. We obtain its deformed shape by minimising the sum of the energies of the individual cells subject to this imposed compression, as described in the Supplemental Material~\footnote{See Supplemental Material at [url to be inserted], which includes Refs.~\cite{haas19,scipy}, for (i) a description of the discrete vertex model and its numerical implementation, (ii)~a more detailed summary of the corresponding continuum limit, (iii) the derivation of the modified continuum equations for triangular cells, and (iv) details of the asymptotic calculations near the unbuckling bifurcation and further comparisons of the numerical and asymptotic results for the continuum model.}. \nocite{haas19,scipy} 

At a critical compression $D=D_\ast$, the monolayer buckles~\figref{fig2}{b}. As $D$ increases beyond $D_\ast$, the buckling amplitude increases until cells in the crest and trough of the buckled shape become triangular at $D=D_\triangle$. For $D>D_\triangle$, fans of triangular cells~\figref{fig2}{c} expand from the crests and troughs. For still larger compressions, we find, surprisingly, two different shape outcomes at the same compression: There is a critical value $r_\text{bif}$ of the aspect ratio $r$ of the undeformed monolayer~\figrefi{fig2}{a} such that, if $r>r_\text{bif}$, the buckling amplitude continues to increase~\figref{fig2}{d} to form tight folds reminiscent of the cephalic furrow~\figref{fig1}{a} or the ectopic folds~\cite{vellutini23}. However if $r<r_\text{bif}$, the amplitude decreases~\figref{fig2}{e}. We will call this ``unbuckling''.

\begin{figure*}
\centering 
\includegraphics[width=17.8cm]{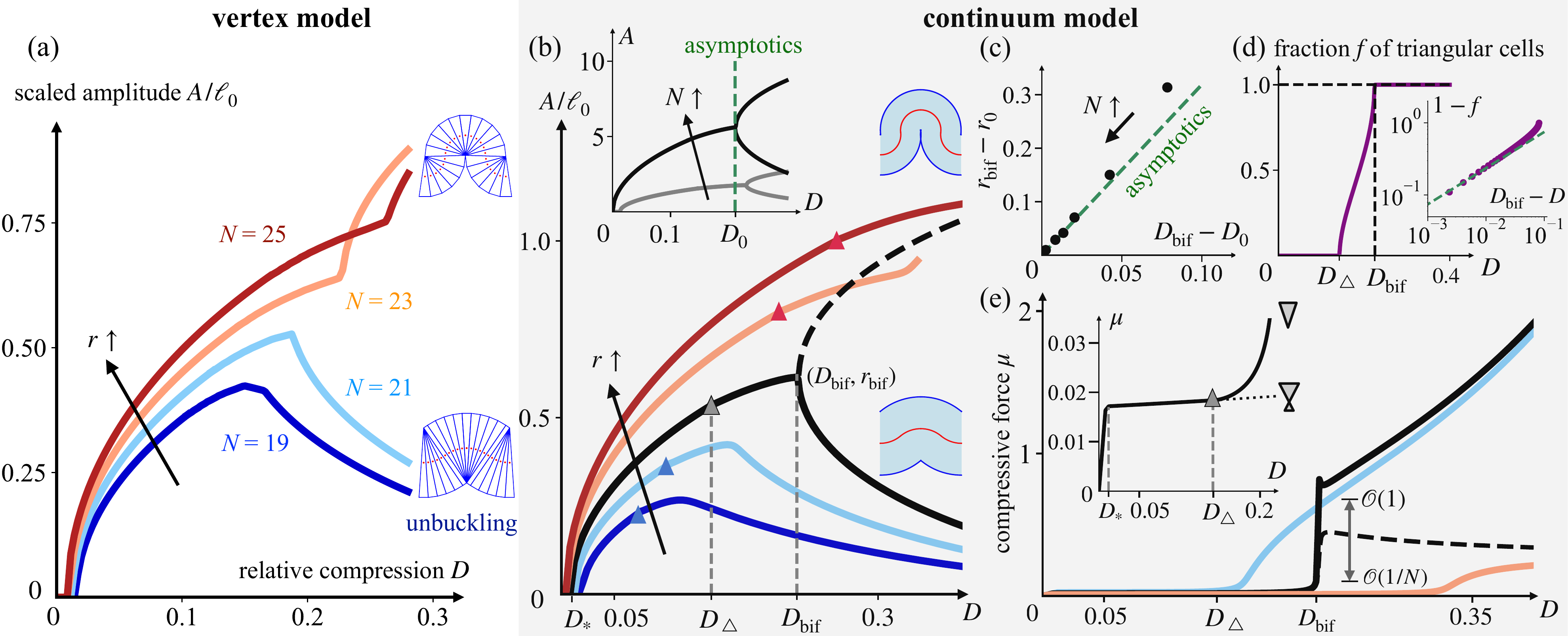}
\caption{Unbuckling bifurcation. (a)~Plot of the buckling amplitude $A$ against relative compression $D$ in the vertex model and for different cell numbers $N$, showing sharp unbuckling for small enough $N$. Parameter value for numerical calculations: $\ell_0=3$. (b)~Corresponding plot for the continuum model, showing the unbuckling transition as the aspect ratio $r$ of the monolayer increases. A critical branch (black lines) at $r=r_\text{bif}$ separates buckling and unbuckling behaviour, bifurcating at $D=D_\text{bif}$. The upper branch is dashed to indicate steric self-intersection for $D>D_\text{bif}$ on this branch. The critical values $D_\ast,D_\triangle$ are highlighted on the $D$-axis for the critical branch $r=r_\text{bif}$. Triangles on branches mark $D=D_\triangle$. Parameter value: $\smash{\ell_0=\sqrt{10}}$. Inset: behaviour of the critical branch $r=r_\text{bif}$ as $N$ and $\ell_0$ increase, showing that $D_\text{bif}\to D_0$; the dashed line shows the prediction $D_0\approx 0.21$ from asymptotic calculations. (c)~Convergence $D_\text{bif}\to D_0$, $r_\text{bif}\to r_0$ as $N$ and $\ell_0$ increase: Numerical results (dot markers) match the asymptotic prediction (dashed line). (d)~Plot of the fraction $f$ of triangular cells against $D$ for the critical branch $r=r_\text{bif}$ in panel (b), rising from $0$ for $D>D_\triangle$ and approaching $1$ as $D$ approaches $D_\text{bif}$. Inset: plot of $1-f$ against $D_\text{bif}-D>0$, confirming that the numerical results (dot markers) match the asymptotic calculations (dashed line). (e)~Plot of the compressive force $\mu$ against relative compression $D$ for the critical branches $r=r_\text{bif}$ in panel (b), showing a large increase near the bifurcation at $D=D_\text{bif}$. The vertical arrow indicates the asymptotic orders of magnitude over which the monolayer stiffens near $D=D_\text{bif}$. Inset: same plot, zooming in on smaller $\mu$ and $D$ and showing the decrease of the apparent stiffness $\partial\mu/\partial D$ of the monolayer associated with the buckling at $D=D_\ast$, and its increase for $D>D_\triangle$ (triangular shape inset); the dotted line shows the slower increase of $\mu$ if the geometric singularity is ignored (self-intersecting shape inset).}
\label{fig3}
\end{figure*}

To understand the mechanical basis of this unbuckling, we consider the continuum limit of an epithelium of ${N\gg 1}$ cells~\cite{haas19}. Before compression, the cells are rectangular, of height ${\ell_0=\sqrt{2\Gamma/\Gamma_\ell}}$ and width $1/\ell_0$ from minimising Eq.~\eqref{eq:sce}. Thus $2r=(N/\ell_0)/\ell_0=N/\ell_0^2$. Let $s$ be arclength along the undeformed midline, so that $\sigma=s/r\ell_0\in[0,2]$. In the limit $\ell_0\gg1$ of columnar cells, and assuming ${D<D_\triangle}$ and a symmetric buckled shape, the Euler--Lagrange equation for the tangent angle $\psi$ of the deformed midline~\figref{fig1}{b} is~\cite{haas19}
\begin{align}
\dfrac{\mathrm{d}^2\psi}{\mathrm{d}\sigma^2}+ \frac{\mu N^2\sin{\psi} }{1+\mu \cos\psi}\left[1-\frac{1}{2N^2}\left(\dfrac{\mathrm{d}\psi}{\mathrm{d}\sigma}\right)^2\right]=0,
\end{align}
subject to the boundary conditions ${\psi(0)=\psi(1)=0}$. Here, $\mu$ is the compressive force, i.e., the Lagrange multiplier that imposes the compression of the monolayer,\begin{subequations}
\begin{align}
\int_0^1\cos{\psi \left[1+\frac{1}{2N^2}\left(\dfrac{\mathrm{d}\psi}{\mathrm{d}\sigma}\right)^2\right]\mathrm{d}\sigma} = \Lambdai(1-D),
\end{align}
in which $\Lambdai$ is the stretch of its lateral cell sides, which is the same for all cells for them to match up. The Supplemental Material~\cite{Note2} shows that $\Lambdai$ is determined by
\begin{align}
\int_0^1{\frac{1}{2N^2}\left(\dfrac{\mathrm{d}\psi}{\mathrm{d}\sigma}\right)^2\mathrm{d}\sigma}=\Lambdai^2-1-\mu\Lambdai(1-D).
\end{align}
\end{subequations}
There, we show further how to modify these governing equations to describe the buckled shapes with triangular cells for $D>D_\triangle$ in this continuum framework, too. Solving these continuum equations recapitulates the behaviour of the discrete model: The initially flat monolayer~\figref{fig2}{f} buckles for $D>D_\ast$~\figref{fig2}{g}. The triangular cells that arise for $D>D_\triangle$ appear as kinks in the buckled continuum shapes~\figref{fig2}{h}. Finally, unbuckling also appears in the continuum model for $D>D_\text{bif}$: At the same compression, the buckling amplitude increases for $r>r_\text{bif}$~\figref{fig2}{i}, but decreases for $r<r_\text{bif}$~\figref{fig2}{j}. 

To quantify the unbuckling transition in our minimal vertex model, we plot,  in \textfigref{fig3}{a}, the buckling amplitude $A$ against the relative compression $D$ for different $N$ and at constant $\ell_0$ (i.e., at constant surface tensions). We observe a sharp decrease of $A$ with $D$ for sufficiently small $N$ that signals unbuckling. We find the same behaviour in the continuum model beyond $D=D_\triangle$, as, equivalently, $r=N/2\ell_0^2$ increases at constant $\ell_0$~\figref{fig3}{b}. A critical branch with $r=r_\text{bif}$ separates the unbuckling branches from those on which $A$ continues to increase. This critical branch bifurcates for ${D>D_\text{bif}}$~\figref{fig3}{b}. Tracking it numerically~\figrefi{fig3}{b} as $N$ and $\ell_0$ increase, we find that $D_\text{bif}\to D_0\approx 0.21$, $r_\text{bif}\to r_0\approx 1.17$ in the continuum model. We are left to understand this bifurcation.

The key insight comes from the plot of the fraction~$f$ of triangular cells on the critical branch $r=r_\text{bif}$ in \textfigref{fig3}{d}: As expected, $f$ increases from $0$ for $D>D_\triangle$, but its approach to $1$ coincides with $D=D_\text{bif}$, and it remains very close to $1$ for $D>D_\text{bif}$. This suggests that shapes in which (almost) all cells have become triangular underlie the unbuckling bifurcation. We will show that this determines the bifurcation geometrically.

\begin{figure}[t]
\centering 
\includegraphics[width=8.6cm]{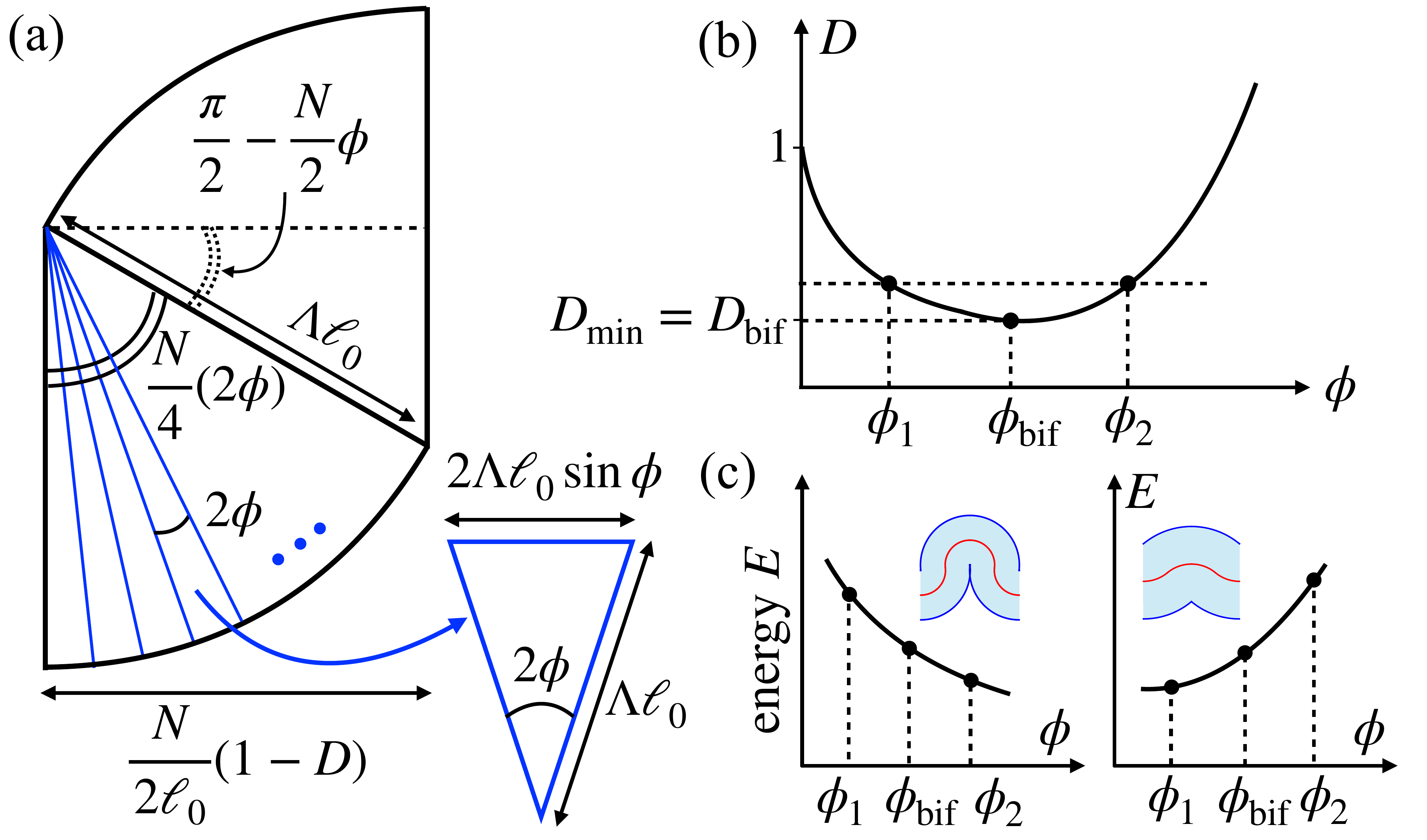}
\caption{Mechanism of unbuckling. (a)~Buckled shapes with all cells triangular: Each half of the buckled shape consists of two matched fans of $N/4$ triangular cells each. Inset: isosceles triangular cell with base angle $2\phi$, lateral side $\Lambdai\ell_0$. (b)~Plot of the relative compression $D$ against the semi-angle $\phi$, with minimum $D_\text{min}=D_\text{bif}$ at $\phi=\phi_\text{bif}$. For $D\gtrsim D_\text{bif}$, there are two solutions, $\phi_1<\phi_\text{bif}$, $\phi_2>\phi_\text{bif}$. (c)~Plots of the energy $E$ against $\phi$: If $\partial E/\partial\phi(\phi_\text{bif})<0$, then $\phi_2$ is selected (buckling); if $\partial E/\partial\phi(\phi_\text{bif})>0$, then $\phi_1$ is selected (unbuckling).}
\label{fig4}
\end{figure}

Indeed, each half of buckled shapes of this kind~\figref{fig4}{a} has width $(N/2\ell_0)(1-D)$, and consists of two matched fans of $N/4$ triangular cells each. These cells have lateral sides $\Lambdai\ell_0$ and base angle $2\phi$~\figrefi{fig4}{a}. This means that each fan has base angle $(N/4)(2\phi)$, so ${(\Lambdai\ell_0)\cos{(\pi/2-N\phi/2)}=(N/2\ell_0)(1-D)}$ \figref{fig4}{a}, implying ${D=1-(\Lambdai/r)\sin{(N\phi/2)}}$ on recalling ${r=N/2\ell_0^2}$. The isosceles cells have base $2\Lambdai\ell_0\sin{\phi}$~\figrefi{fig4}{a}, so the buckled shape has energy ${E=N[\Gamma_\ell(\Lambdai\ell_0)+\Gamma(2\Lambdai\ell_0\sin{\phi})]}$. Using cell area conservation $(\Lambdai\ell_0)^2\sin{2\phi}=2$ and ${\ell_0^2=2\Gamma/\Gamma_\ell=N/2r}$ to eliminate $\ell_0,\Lambdai$, we then obtain

\phantom{.}\vspace{-16pt}
\begin{align}
&D=1-\dfrac{2\sin{(N\phi/2)}}{\sqrt{r N\sin{2\phi}}},&&\dfrac{E}{\sqrt{2}N\Gamma_\ell}=\dfrac{2r+N\sin{\phi}}{2r\sqrt{\sin{2\phi}}}.\label{eq:DE}
\end{align}
The first of these has a solution if and only if $D\geqslant D_\text{min}$, for some $D_\text{min}$~\figref{fig4}{b}. Since the bifurcation coincides with the first appearance of buckled shapes in which all cells are triangular~\figref{fig3}{d}, $D_\text{min}=D_\text{bif}$, and we denote by $\phi_\text{bif}$ the corresponding semi-angle. For $D$ just above $D_\text{bif}$, this equation has two solutions, $\phi_1<\phi_\text{bif}$ and ${\phi_2>\phi_\text{bif}}$, respectively decreasing and increasing as $D$ increases \figref{fig4}{b}, and hence corresponding to unbuckling and buckling, respectively. Now, if $\partial E/\partial\phi(\phi_\text{bif})<0$, then $E(\phi_1)>E(\phi_2)$~\figref{fig4}{c} and the buckling branch is energetically favourable. Similarly, if $\partial E/\partial\phi(\phi_\text{bif})>0$, then $E(\phi_2)<E(\phi_1)$, leading to unbuckling. All of this shows that $\partial D/\partial\phi=\partial E/\partial\phi=0$ at $\phi=\phi_\text{bif}$, so
\begin{align}
&\tan{\dfrac{N\phi_\text{bif}}{2}}=\dfrac{N}{2}\tan{2\phi_\text{bif}},&& \sin{\phi_\text{bif}}=\dfrac{2r_\text{bif}}{N}\cos{\phi_\text{bif}}.\label{eq:bif}
\end{align}
For $N\gg 1$, $r_\text{bif}\to r_0$, $D_\text{bif}\to D_0$, and $\phi_\text{bif}\to\phi_0\ll 1$. The second of Eqs.~\eqref{eq:bif} thus yields $\phi_0=2r_0/N$, whence, from the first of Eqs.~\eqref{eq:bif} and of Eqs.~\eqref{eq:DE},
\begin{align}
&\tan{r_0}=2r_0,&&D_0=1-\dfrac{\sin{r_0}}{r_0}.\label{eq:bifsol}
\end{align}
This implies $r_0\approx 1.165$ and hence $D_0\approx0.212$, in excellent agreement with the numerical solutions of the full continuum equations~\figrefi{fig3}{b}.

We can confirm the existence of the unbuckling bifurcation more formally by asymptotic solution of these equations for $f\approx 1$, near the bifurcation point ${D=D_0}$, $r=r_0$. We thus write ${D\!=\!D_0\!+\!\delta\xi^2}$, $r\!=\!r_0\!+\!\rho\xi^2$, where $\xi\!=\!2/N\!\ll\! 1$ is the small parameter for the expansion, and the control parameters $\delta,\rho$ are of order $\mathcal{O}(1)$, with the bifurcation occurring at $\delta=\delta_\text{bif},\rho=\rho_\text{bif}$. We relegate the unedifying details of these calculations to the Supplemental Material~\cite{Note2}, and summarise the results here. First, the fraction $f$ of triangular cells admits precisely two expansions: (i)~${f=1\!-\!f_1\xi\!+\!\mathcal{O}\bigl(\xi^2\bigr)}$, (ii) ${f=1\!-\!f_2\xi^2\!+\!\mathcal{O}\bigl(\xi^3\bigr)}$, where $f_1,f_2>0$. For both expansions, we recover Eqs.~\eqref{eq:bifsol}. The first expansion then leads to $\smash{f_1^2=\bigl(3\rho-r_0^3-3r_0\delta\sec{r_0}\bigr)\big/\bigl(2r_0^3\bigr)}$, which is well-defined only for $\smash{\delta<\bigl(\rho-r_0^3/3\bigr)\cos{r_0}}$. This gives $\smash{\delta_\text{bif}=\bigl(\rho_\text{bif}-r_0^3/3\bigr)\cos{r_0}}$. As $\delta\to\delta_\text{bif}$, $f_1\to 0$, indicating a transition from the first expansion to the second. For this, we find $f_2=2r_0^2$, but now $\Lambdai=1+\Lambdai_1\xi+\mathcal{O}\bigl(\xi^2\bigr)$, where $\smash{\Lambdai_1^2=\bigl(r_0^3-3\rho+3r_0\delta\sec{r_0}\bigr)\big/\bigl[3r_0\bigl(4r_0^2-1\bigr)\bigr]}$, well-defined only for $\smash{\delta>\bigl(\rho-r_0^3/3\bigr)\cos{r_0}}$, so expansions~(i) and~(ii) cover all possible values of the control parameters $\delta,\rho$. Moreover, the two possible solutions ${\Lambdai_1\gtrless 0}$ herald the two solution branches of the bifurcation. Our calculations~\cite{Note2} finally yield ${r_\text{bif}-r_0\approx 3.188(D_\text{bif}-D_0)}$ and $1-f\approx 2.367(D_\text{bif}-D)^{1/2}$ for $D<D_\text{bif}$. These are in excellent agreement with the numerical solutions~[\textfigref{fig3}{c} and \textfigref{fig3}{d}, inset], as are other results in the Supplemental Material~\cite{Note2}.

This mathematical exercise finds another meaning in the relation between the compressive force $\mu$ acting on the monolayer and its compression $D$~\figref{fig3}{e}. As $D$ increases from~$0$, $\mu$ increases sharply initially~\figrefi{fig3}{e}, but then increases more slowly. This reduction of the effective stiffness $\partial\mu/\partial D$ is associated with the initial buckling of the epithelium at $D=D_\ast$. With the appearance of triangular cells at $D=D_\triangle$, the stiffness increases once more~\figrefi{fig3}{e}. This increase is tiny, however, compared to the drastic increase of $\mu$ near $D=D_\text{bif}$ \figref{fig3}{e}. The latter is in fact asymptotically large: The calculations in the Supplemental Material~\cite{Note2} show that $\mu=\mu_1\xi+\mathcal{O}\bigl(\xi^2\bigr)$ for $\delta<\delta_\text{bif}$, with $\mu_1$ diverging as $\delta\to\delta_\text{bif}$ to transition to $\mu=\mu_0+\mathcal{O}(\xi)$ for $\delta>\delta_\text{bif}$. Importantly, the stiffening is not therefore the result of the appearance of individual triangular cells (that causes the slight stiffening near $D=D_\triangle$ mentioned above) but the result of the unbuckling bifurcation, i.e., the collective effect of all cells becoming triangular. It is not the result of steric interactions between different cells in contact either~\footnote{The critical buckling branch at $r=r_\text{bif}$ [dashed lines in Figs.~\figrefp{fig3}{b),(e}] self-intersects sterically for all ${D>D_\text{bif}}$. The snapping instability that is implied by $\partial\mu/\partial D<0$ on this branch~\figref{fig3}{e} is therefore unphysical.}. Steric interactions do arise in the cephalic furrow~\figref{fig1}{a} and in the ectopic folds~\cite{vellutini23} that correspond to the ``buckling'' branch of the bifurcation. They are not described by our continuum model, but importantly, they only arise for $D>D_\text{bif}$ where we expect them to continue stiffening the epithelium.

In summary, we have studied the behaviour of epithelial monolayers under compression in detail in a minimal vertex model and in the corresponding continuum model. Our combined numerical and analytical results have uncovered an ``unbuckling'' bifurcation. The asymptotically large increase of the stiffness of the monolayer associated with this bifurcation links it to the ability of tight tissue folds such as the cephalic furrow~[\citenum{vellutini23}, \citenum{dey23}, \textfigref{fig1}{a}] to absorb compressive stresses. This ascribes a biomechanical function to the unbuckling transition.

Our unbuckling bifurcation lacks a direct analogue in classical continuum instabilities because it results from the geometric singularity of cell constriction of the vertex model. Nevertheless, the cuspy shapes close to the bifurcation are reminiscent of the buckled shapes of thick compressed hyperelastic columns~\cite{chen20}. It is also known that the amplitude of the creasing instability~\cite{hong09,hohlfeld11} of a compressed elastic block can decrease with increasing compression for a strain-stiffening material~\cite{jin15}, which recalls the unbuckling described here.

Meanwhile, generalisations of our minimal vertex model~\cite{Note1} undergo a wrinkling instability at sufficiently large apico-basal tension difference~\cite{matej13,matej23}, and an ``unwrinkling'' transition at even larger apicobasal tension difference~\cite{matej13}. Future work, starting by extending their continuum limit~\cite{matej23} to a large-deformation theory, will need to determine the mechanism underlying this unwrinkling and its relation to our unbuckling bifurcation.

\begin{acknowledgments}
We thank Marija Krstic for discussions and preliminary computations at an early stage of this work. We also thank Bruno Vellutini for providing the image in \textfigref{fig1}{a} and for comments on the manuscript, and Matt Bovyn and Mike Staddon for further comments on the manuscript. We gratefully acknowledge funding from the Max Planck Society.
\end{acknowledgments}
\bibliography{main}
\end{document}


\renewcommand{\theequation}{S\arabic{equation}}
\renewcommand{\thefigure}{S\arabic{figure}}
\renewcommand{\thepage}{S\arabic{page}}
\renewcommand{\thetable}{S}
\title{(Un)buckling mechanics of epithelial monolayers under compression\\\textbullet{} Supplemental Material \textbullet{}}

\author{Chandraniva \surname{Guha Ray}}
\affiliation{Max Planck Institute for the Physics of Complex Systems, N\"othnitzer Stra\ss e 38, 01187 Dresden, Germany}
\affiliation{\smash{Max Planck Institute of Molecular Cell Biology and Genetics, Pfotenhauerstra\ss e 108, 01307 Dresden, Germany}}
\affiliation{Center for Systems Biology Dresden, Pfotenhauerstra\ss e 108, 01307 Dresden, Germany}
\author{Pierre A. Haas}
\affiliation{Max Planck Institute for the Physics of Complex Systems, N\"othnitzer Stra\ss e 38, 01187 Dresden, Germany}
\affiliation{\smash{Max Planck Institute of Molecular Cell Biology and Genetics, Pfotenhauerstra\ss e 108, 01307 Dresden, Germany}}
\affiliation{Center for Systems Biology Dresden, Pfotenhauerstra\ss e 108, 01307 Dresden, Germany}\maketitle

This Supplemental Material divides into four parts: We first describe the discrete vertex model along with its numerical implementation. We then summarise the governing equations of the continuum theory. We go on to discuss how to modify these equations to take into account the presence of triangular cells. Finally, we provide details of the asymptotic calculations near the ``unbuckling'' bifurcation, along with further comparisons between the numerical and asymptotic results.

\section{Discrete vertex model}
\begin{figure}[b]
\centering 
\includegraphics[width=8.6cm]{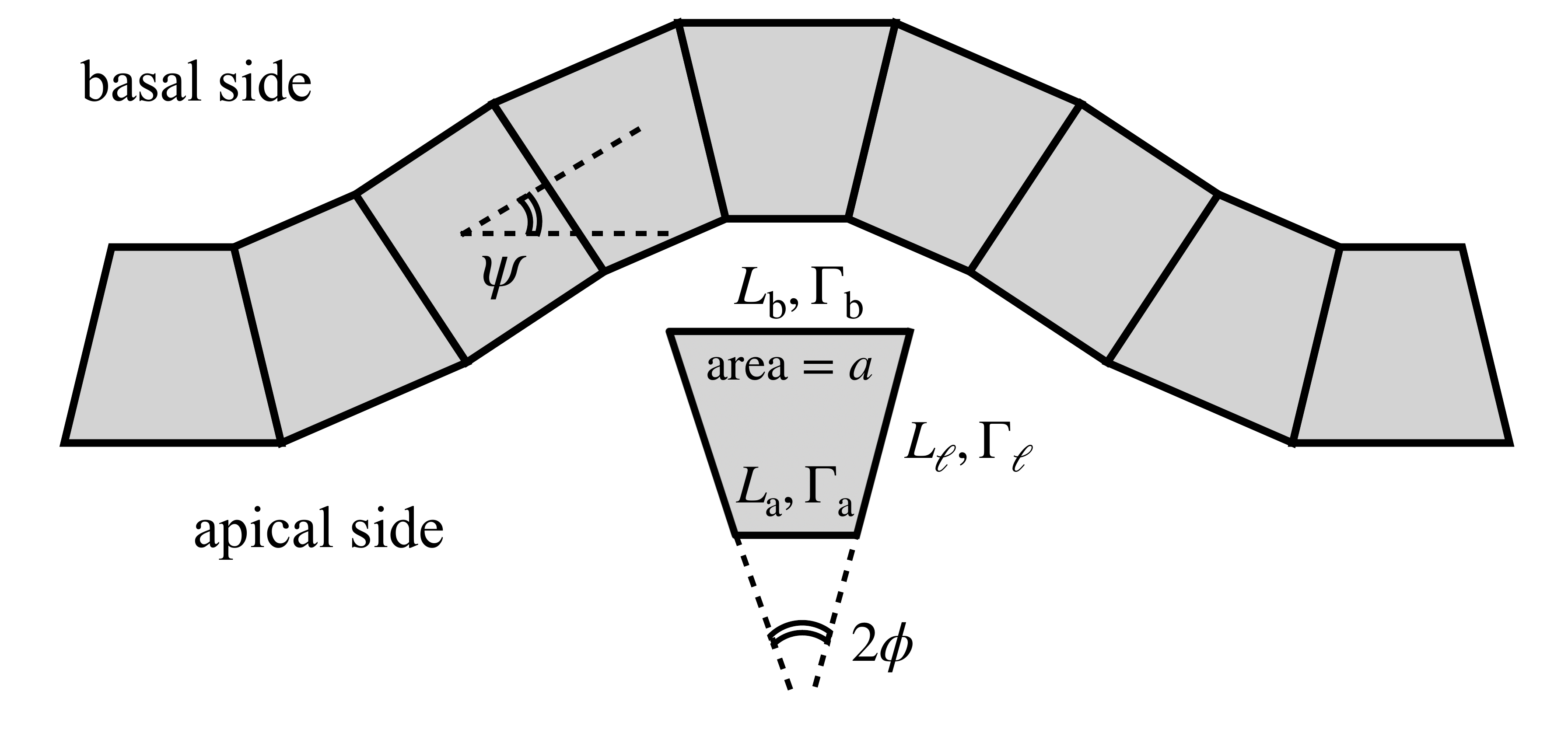}
\caption{Model epithelium of $N$ isosceles trapezoidal cells of apical, basal, and lateral sides $L_{\text{a}}$, $L_{\text{b}}$, $L_\ell$, respectively, and corresponding surface tensions $\Gamma_{\text{a}}$, $\Gamma_{\text{b}}$, $\Gamma_\ell$. The lateral cell sides are at an angle $2\phi$ to each other, and the cell midline makes an angle $\psi$ with the vertical. Each cell has area $a$.}\label{figS1}
\label{fig:model}
\end{figure}
We analyse a minimal vertex model in which the energy of a single, isosceles trapezoidal cell~\wholefigref{figS1} is
\begin{subequations}
\begin{align}
E=\Gamma_{\mathrm{a}}L_{\mathrm{a}}+\Gamma_{\mathrm{b}}L_{\mathrm{b}}+\Gamma_\ell L_\ell,
\end{align}
where $L_{\mathrm{a}},L_{\mathrm{b}},L_\ell$ are the respective lengths of its apical, basal, and lateral sides, and, $\Gamma_{\mathrm{a}},\Gamma_{\mathrm{b}},\Gamma_\ell$ are the corresponding surface tensions, subject to the constraint of fixed cell area $a$. We define $\psi$ to be the angle made by the midline of the cell with the horizontal and $2\phi$ to be the angle made by its lateral sides with each other~\wholefigref{figS1}.
\subsection{Nondimensionalisation}
We will henceforth take $\Gamma_{\mathrm{a}}=\Gamma_{\mathrm{b}}\equiv\Gamma$ as in the main text, and we will, as in Ref.~\cite{haas19}, nondimensionalise lengths with~ $\smash{\sqrt{a}}$ and energies with $\smash{\Gamma_\ell/\sqrt{a}}$. 
Thus the nondimensional length of the lateral sides is $\lambda = \smash{L_\ell/\sqrt{a}}$. With this nondimensionalisation and similarly to Ref.~\cite{haas19}, the single-cell energy becomes
\begin{align}
e=\ell_0\left(\dfrac{\sec{\phi}}{\Lambda}+\Lambda\right),
\end{align}
\end{subequations}
where $\ell_0=\sqrt{2\Gamma/\Gamma_\ell}$ is the cell height minimising the energy of a single cell in its undeformed, rectangular configuration, and $\Lambda= \lambda/\ell_0$ is the stretch of the lateral sides of the deformed monolayer. 
\subsection{Numerical implementation}
In the main text, we consider the buckling of an epithelial monolayer of $N$ cells with clamped ends due to an imposed relative compression $D$. The total energy of the monolayer is thus
\begin{align}
\mathcal{E}=\sum_{i=1}^N \ell_0\left(\dfrac{\sec{\phi_i}}{\Lambda}+\Lambda\right),
\end{align}
where $2\phi_i$ is now the angle that the lateral sides of cell $i$ make with each other, for $i=1,2,\dots,N$. The stretch $\Lambda$ of the lateral sides is a constant (i.e., the same for all cells) because the adjacent cell sides have to match up. 

We will seek symmetrically buckled solutions, which reduces the problem to minimising the energy of half of the monolayer,
\begin{align}
\dfrac{\mathcal{E}}{2\ell_0}=\left\{\begin{array}{l}
\displaystyle\sum_{i=1}^n{\left(\dfrac{\sec{\phi_i}}{\Lambda}+\Lambda\right)}\qquad\text{if $N=2n$ is even};\\
\displaystyle\sum_{i=1}^n\left(\dfrac{\sec{\phi_i}}{\Lambda}+\Lambda\right)+\dfrac{1}{2}\left(\dfrac{\sec{\phi_n}}{\Lambda}+\Lambda\right)\\
\phantom{.}\hspace{29.6mm}\text{if $N=2n-1$ is odd}.\label{eq:halfe}
\end{array}\right.
\end{align}
Let $\psi_i$ denote the angle that cell $i$ makes with the horizontal, for $i=1,2,\dots,N$. The clamped boundary conditions imply $\psi_1=0$. If $N=2n-1$ is odd, the assumed symmetry of the buckled shape further implies $\psi_n = 0$. This condition can be imposed using the recursive geometric relation~\cite{haas19}
\begin{align}
\psi_i = \psi_{i-1} + \phi_i + \phi_{i-1}.
\end{align}
Let $x_i$ denote the horizontal position of the centre of cell $i$, for $i=1,2,\dots,N$, which, with $x_1=0$, satisfies the recursive relation~\cite{haas19}
\begin{align}
x_i = x_{i-1} + \frac{\sec{\phi_i}\cos{\psi_i} + \sec{\phi_{i-1}} \cos{\psi_{i-1}}}{2\ell_0\Lambda}.
\end{align}
The imposed compression of the monolayer then yields the constraint
\begin{align}
x_n=\!\left\{\begin{array}{ll}
\dfrac{n-1}{\ell_0}(1-D)&\text{if $N=2n-1$ is odd};\\
\dfrac{2n-1}{2\ell_0}(1-D)-\dfrac{\sec{\phi_n}\cos{\psi_n}}{2\ell_0\Lambda}&\text{if $N=2n$ is even}.
\end{array}\right.
\end{align}
We minimise the energy in Eq.~\eqref{eq:halfe} subject to these constraints over $\{\phi_1,\dots,\phi_n\}$ to find the buckled shape of the monolayer using the \textsc{python3} optimisation routine \texttt{scipy.optimize}~\cite{scipy}. This scheme works as long as the compression is small enough for the cells not to self-intersect. 

The condition to avoid self-intersection of the isosceles trapezoidal cells is~\cite{haas19}
\begin{align}
& \lambda^2 \leq \dfrac{2}{\left|\sin{2\phi_i}\right|}\quad\Longrightarrow\quad |\phi_i| \leq \frac{1}{2}\sin^{-1}{\left(\dfrac{2}{\lambda^2}\right)} \equiv \phi_{\rm max}.
\end{align}
We implement this new inequality constraint by minimisation over the new variables $\Phi_1,\dots,\Phi_n$ such that
\begin{align}
\phi_i = \phi_{\rm max}\sin{\Phi_i}.
\end{align}

\section{Continuum model}
To describe the continuum limit of this buckling problem, we consider an epithelial monolayer of undeformed length $2\Sigma\gg \ell_0^{-1}$ in its flat configuration, clamped at its ends, under an imposed relative compression $D$. By definition, the number of cells in the monolayer is $N=2\Sigma/\ell_0^{-1}$. As in the numerical implementation of the discrete problem, we restrict to symmetrically buckled shapes, i.e., to the range $0\leq s\leq \Sigma$ of the nondimensional arclength in the undeformed configuration. The Lagrangian describing this continuum limit for columnar cells ($\ell_0\gg 1$) has been derived in Ref.~\cite{haas19}. Introducing $\sigma=s/\Sigma$, this Lagrangian is
\begin{align}
 \mathcal{L} &= \int_0^1{\left[\Lambda + \frac{1}{\Lambda}+ \frac{\dot{\psi}^2 }{8 \Lambda \Xi^2}\right]\mathrm{d}\sigma}  \nonumber\\
   &\quad+\mu \left[\int_0^1{\frac{ \cos{\psi}}{\Lambda}\left( 1+\frac{\dot{\psi}^2}{8\Xi^2}\right)\mathrm{d}\sigma}-(1-D)\right]\label{eq:L}
\end{align}
to leading order in an expansion in $\Xi=\ell_0\Sigma=N/2\gg 1$, where dots denote differentiation with respect to $\sigma$, where $\psi(\sigma)$ is now the continuous tangent angle to the midline of the epithelium and $\mu$ is the Lagrange multiplier that imposes its prescribed compression via the integral condition
\begin{equation}
\int_0^1\cos{\psi \left( 1+\frac{\dot{\psi}^2}{8\Xi^2} \right)\mathrm{d}\sigma} = \Lambda(1-D).
\end{equation}
The clamping of the ends of the epithelium and the assumed symmetry of the buckled shape yield the boundary conditions
\begin{align}
\psi(0)=\psi(1)=0.\label{eq:bc}
\end{align}
The variation of the Lagrangian~\eqref{eq:L} is
\begin{widetext}
\vspace{-12pt}
\begin{align}
\delta \mathcal{L}&=\delta\mu\left[\int_0^1{\frac{\cos\psi}{\Lambda} \left(1+\frac{\dot{\psi}^2}{8\Xi^2}\right)\mathrm{d}\sigma}-(1-D)\right] + \delta \Lambda \left[1-\frac{1}{\Lambda^2}-\frac{1}{8\Lambda^2\Xi^2}\int_0^1{\dot{\psi}^2\mathrm{d}\sigma} - \frac{\mu}{\Lambda^2}\int_0^1{\cos\psi\left(1+\frac{\dot{\psi}^2}{8\Xi^2} \right)\mathrm{d}\sigma}\right]\nonumber\\
&\quad+\hspace{-2.5cm}\underbrace{\int_0^1{\left\{\delta \psi  \left[-\frac{\mu\sin\psi}{\Lambda} \left(1+\frac{\dot{\psi}^2}{8\Xi^2}\right) \right] +\delta \dot{\psi} \left(\frac{\dot{\psi}}{4\Lambda\Xi^2} + \frac{\mu \dot{\psi}}{4\Lambda\Xi^2}  \cos{\psi} \right)\right\}\mathrm{d}\sigma}.}_{\displaystyle\phantom{.}\hspace{2.5cm}\left\llbracket \frac{\dot{\psi}}{4\Lambda\Xi^2}(1+\mu \cos\psi)\,\delta\psi\right\rrbracket_0^1-\int_0^1{\left[\dfrac{\ddot{\psi}}{4\Lambda\Xi^2}(1+\mu\cos{\psi})+\dfrac{\mu\sin{\psi}}{\Lambda}\left(1-\dfrac{\dot{\psi}^2}{8\Xi^2}\right)\right]\delta\psi\,\mathrm{d}\sigma}} 
\end{align}
\end{widetext}
Noting that the boundary terms in the last line above vanish because of the boundary conditions~\eqref{eq:bc}, we read off the governing equation
\begin{align}
\ddot{\psi} + \frac{4 \mu \Xi^2\sin{\psi} }{1+\mu \cos\psi}\left(1-\frac{\dot{\psi}^2}{8\Xi^2}\right)=0,\label{eq:gov}
\end{align}
subject again to the boundary conditions~\eqref{eq:bc} and subject to the integral conditions
\begin{subequations}
\begin{align}
&\int_0^1{\frac{\cos\psi}{\Lambda} \left(1+\frac{\dot{\psi}^2}{8\Xi^2}\right)\mathrm{d}\sigma}=1-D,\\
&\int_0^1{\frac{\dot{\psi}^2}{8\Xi^2}\,\mathrm{d}\sigma}=\Lambda^2-1-\mu\Lambda(1-D),
\end{align}
\end{subequations}
where we have used the first condition to simplify the second one, which was not derived in Ref.~\cite{haas19}. The two boundary conditions~\eqref{eq:bc}, along with these integral conditions, determine the shape $\psi(\sigma)$ of the midline the epithelium via the second-order differential equation~\eqref{eq:gov} and the two unknown constants $\Lambda$ and $\mu$, i.e., the stretch of its lateral sides and the compressive force on the monolayer~\footnote{The shape $\bigl(x(\sigma),y(\sigma)\bigr)$ of the deformed midline is determined, at our order of approximation~\cite{haas19} and from the solution for the tangent angle $\psi(\sigma)$, by the differential equations
\begin{align*}
\dot{x}&=\dfrac{\Xi}{\Lambda\ell_0}\left[\left(1 + \frac{\dot{\psi}^2}{24 \Xi^2}\right)\cos{\psi}-\frac{\ddot{\psi}}{12 \Xi^2}
\sin{\psi}\right],\\
\dot{y}&=\dfrac{\Xi}{\Lambda\ell_0}\left[\left(1 + \frac{\dot{\psi}^2}{24 \Xi^2}\right)\sin{\psi}+\frac{\ddot{\psi}}{12 \Xi^2}\cos{\psi}\right].
\end{align*}
The surfaces $\bigl(x_+(\sigma),y_+(\sigma)\bigr)$, $\bigl(x_-(\sigma),y_-(\sigma)\bigr)$ of the monolayer satisfy~\cite{Note2}
\begin{align*}
x_\pm(\sigma)&=x_\pm(\sigma)\mp\dfrac{\ell_0}{2}\Lambda\cos{\phi(\sigma)}\sin{\psi(\sigma)},\\
y_\pm(\sigma)&=y_\pm(\sigma)\pm\dfrac{\ell_0}{2}\Lambda\cos{\phi(\sigma)}\cos{\psi(\sigma)},
\end{align*}
where, at this order of approximation, $\phi(\sigma)=\dot{\psi}(\sigma)/2\Xi$~\cite{haas19}. We integrate these equations numerically to obtain the buckled monolayer shapes shown in the main text. It is straightforward to extend this to the case of triangular cells, by splitting the range of integration into $0\leq\sigma\leq\sigma_\ast$ and $\sigma_\ast\leq\sigma\leq 1/2$.
}.

\vspace{-8pt}
\section{Modified governing equations for triangular cells}
Our model breaks down under high relative compression when one side of the isosceles trapezoidal cells constricts to the point of vanishing. In order to prevent this self-intersection, we have to impose that the cells cannot constrict further once they have become triangular. We do this by modifying our governing equations to account for triangular cells. 

We recall that the arclength range $0\leq\sigma\leq 1$ represents half of the buckled shape of the monolayer. Because we assume that $\Gamma_{\mathrm{a}}=\Gamma_{\mathrm{b}}$, there is an additional symmetry around ${\sigma=1/2}$, corresponding to a quarter of the total arclength, viz., $\psi(\sigma)=\psi(1-\sigma)$, which, on differentiation, implies $\dot{\psi}(1/2)=0$. In what follows, we will therefore solve the problem on ${0\leq\sigma\leq 1/2}$, subject to $\psi(0)=0$ and $\dot{\psi}(1/2)=0$. 

We expect triangular cells to form first near the troughs and crests of the buckled shapes, where the curvature is largest, i.e. close to $\sigma=0$ (and $\sigma=1$). We therefore divide the range $0\leq\sigma\leq 1/2$ into two subranges,
\begin{enumerate}[label=(\alph{enumi}),widest=2,leftmargin=*,itemsep=-3pt,topsep=3pt]
\item $0\leq \sigma\leq\sigma_\ast$, in which the cells become triangular;
\item $\sigma_\ast\leq\sigma\leq 1/2$, in which the cells are not triangular.
\end{enumerate}
Thus the fraction of triangular cells is $f=2\sigma_\ast$. In the discrete model, $\phi_i=\phi_{\max}$ for triangular cells, so the function $\phi(\sigma)$ representing the semi-angle of the lateral sides in the continuum limit is constant throughout the first range. Since $\dot{\psi}(\sigma)=2\Xi\phi(\sigma)$ at the order of approximation that we are using~\cite{haas19}, it follows that $\dot{\psi}=\Psi(\Lambda;\ell_0,\Xi)$ is also constant throughout this range, whence $\psi(\sigma)=\Psi(\Lambda;\ell_0,\Xi)\sigma$ for ${0\leq\sigma\leq\sigma_\ast}$ using $\psi(0)=0$. 

\subsection{Calculation of $\vec{\Psi(\Lambda;\ell_0,\Xi)}$}
To determine $\Psi(\Lambda;\ell_0,\Xi)$ in the continuum limit, we note that, in a fan of triangular cells, the apical or basal surface of the monolayer constricts to a point. Letting $x_\pm(\sigma)$ denote the horizontal coordinates of the top and bottom surfaces of the monolayer, respectively, it follows that
\begin{align}
\dot{x}_+(0)=0\quad\text{or}\quad\dot{x}_-(0)=0.
\end{align}
Without loss of generality (because of the assumed apicobasal symmetry in our minimal model), we consider buckled shapes with a trough at $\sigma=0$, and thus solve the condition $\dot{x}_+(0)=0$. Let $x(\sigma)$ denote the horizontal coordinate of the midline of the monolayer. Then~\footnote{We note that the expressions for $x_\pm(\sigma),y_\pm(\sigma)$ in Eqs. (42) of Ref.~\cite{haas19} are missing a factor of $1/2$. Consequently, the subsequent results in that paper are for $\ell_0\mapsto 2\ell_0$.}
\begin{align}
x_+(\sigma)=x(\sigma)-\dfrac{\ell_0}{2}\Lambda\cos{\phi(\sigma)}\sin{\psi(\sigma)}.
\end{align}
At the order of approximation that we are using~\cite{haas19},
\begin{align}
\dfrac{\Lambda\ell_0}{\Xi}\dot{x}=\left(1 + \frac{\dot{\psi}^2}{24 \Xi^2}\right)\cos{\psi}-\frac{\ddot{\psi}}{12 \Xi^2}\sin{\psi}.
\end{align}
With $\psi(0)=0$, $\dot{\psi}(0)=\Psi$, $\phi(0)=\dot{\psi}(0)/2\Xi$, it follows that
\begin{align}
\dot{x}_+(0)=\dfrac{\Xi}{\Lambda\ell_0}\left(1+\dfrac{\Psi^2}{24\Xi^2}\right)-\dfrac{\ell_0}{2}\Lambda\Psi\left(1-\dfrac{\Psi^2}{8\Xi^2}\right).
\end{align}
on expanding $\cos{\phi(0)}$ in inverse powers of $\Xi^2$. The condition $\dot{x}_+(0)=0$ thus reduces to the cubic equation
\begin{align}
\frac{1}{\Lambda}- \frac{\ell_0^2\Lambda}{2\Xi}\Psi+\frac{1}{24\Lambda\Xi^2} \Psi^2+\frac{\ell_0^2\Lambda}{16\Xi^3}\Psi^3=0\label{eq:gov04}
\end{align}
\phantom{.}\vspace{-1pt}

\noindent for $\Psi(\Lambda;\ell_0,\Xi)$. We can differentiate this equation with respect to $\Lambda$ to obtain an expression of $\partial\Psi/\partial\Lambda$ that we will need later. We find
\begin{align}
\pdv{\Psi}{\Lambda}  = \dfrac{ \dfrac{1}{\Lambda^2}+\dfrac{\ell_0^2\Psi}{2\Xi}+\dfrac{\Psi^2}{24\Lambda^2\Xi^2} -\dfrac{\ell_0^2\Psi^3}{16\Xi^3}}{-\dfrac{\ell_0^2\Lambda}{2\Xi}+\dfrac{\Psi}{12\Lambda\Xi^2}+\dfrac{3 \ell_0^2\Lambda\Psi^2}{16\Xi^3}}.\label{eq:dpsiL}
\end{align}

\subsection{Modified governing equations}
Using the symmetry around $\sigma=1/2$ discussed above, splitting the range of integration further into $0\leq\sigma\leq\sigma_\ast$ and $\sigma_\ast\leq\sigma\leq 1/2$, and using $\psi(\sigma)=\Psi\sigma$ for $0\leq\sigma\leq\sigma_\ast$, the Lagrangian~\eqref{eq:L} becomes
\begin{widetext}
\vspace{-12pt}
\begin{align}
\mathcal{L} & = 2\int_0^{1/2}{\left(\Lambda + \frac{1}{\Lambda}+ \frac{\dot{\psi}^2 }{8 \Lambda \Xi^2}\right)\mathrm{d}\sigma} +2\mu \left[ \int_0^{1/2}{\frac{ \cos{\psi}}{\Lambda}\left( 1+\frac{\dot{\psi}^2}{8\Xi^2}\right)\mathrm{d}\sigma}-\frac{1-D}{2} \right]\nonumber\\
& = \Lambda + \frac{1}{\Lambda} + \frac{\Psi^2\sigma_\ast}{4 \Lambda \Xi^2}+\dfrac{2\mu}{\Psi\Lambda}\left( 1+\frac{\Psi^2}{8\Xi^2}\right)\sin{(\Psi\sigma_\ast)} + \int_{\sigma_\ast}^{1/2}{\frac{\dot{\psi}^2}{4 \Lambda \Xi^2}\mathrm{d}\sigma}+ 2\mu \left[\int_{\sigma_\ast}^{1/2}{\frac{\cos{\psi}}{\Lambda}\left( 1+\frac{\dot{\psi}^2}{8\Xi^2}\right)\mathrm{d}\sigma}-\frac{1-D}{2}\right].
\end{align}
The variation of the Lagrangian is now
\begin{align}
\delta \mathcal{L}&=2\,\delta\mu\left[  \frac{\sin{(\Psi\sigma_\ast)}}{\Psi\Lambda}\left(1+\frac{\Psi^2}{8\Xi^2}\right) +  \int_{\sigma_\ast}^{1/2}{\frac{\cos{\psi}}{\Lambda}\left( 1+\frac{\dot{\psi}^2}{8\Xi^2}\right)\mathrm{d}\sigma}-\frac{1-D}{2}\right]\nonumber\\
&\quad+\delta \Lambda\left\{1-\frac{1}{\Lambda^2}+2\left(\frac{\Psi}{4\Lambda\Xi^2}\pdv{\Psi}{\Lambda}-\frac{\Psi^2}{8\Lambda^2\Xi^2}\right)\sigma_\ast - \int_{\sigma_\ast}^{1/2}{\frac{\dot{\psi}^2}{4\Lambda^2\Xi^2}\,\mathrm{d}\sigma }+2\mu\left[\frac{\sigma_\ast\cos{(\Psi \sigma_\ast)}}{\Psi\Lambda}\pdv{\Psi}{\Lambda}\left(1+\frac{\Psi^2}{8\Xi^2}\right)\right.\right.\nonumber\\
&\qquad\qquad\left.\left.-\frac{\sin{(\Psi\sigma_\ast)}}{\Psi\Lambda^2}\left(1+\frac{\Psi^2}{8\Xi^2}\right)-\dfrac{\sin{(\Psi\sigma_\ast)}}{\Psi^2\Lambda}\pdv{\Psi}{\Lambda}\left(1-\dfrac{\Psi^2}{8\Xi^2}\right)- \int_{\sigma_\ast}^{1/2}{\frac{\cos \psi}{\Lambda^2}\left(1+\frac{\dot{\psi}^2}{8\Xi^2}\right)\mathrm{d}\sigma}\right]\right\}\nonumber\\
&\quad+2\,\delta \sigma_\ast\left[ \frac{\Psi^2 }{8 \Lambda \Xi^2}  - \frac{\dot{\psi}^2(\sigma_\ast^+) }{8 \Lambda \Xi^2} + \frac{\mu\cos{(\Psi\sigma_\ast)}}{\Lambda}\left(1+\frac{\Psi^2}{8\Xi^2}\right) - \frac{\mu\cos\psi(\sigma_* ^+)}{\Lambda} \left(1+\frac{\dot{\psi}^2(\sigma_* ^+)}{8\Xi^2}\right) \right]\nonumber
\end{align}
\begin{align}
&+2\hspace{-2.7cm}\underbrace{\int_{\sigma_\ast}^{1/2}{\left\{\delta \psi\left[-\frac{\mu\sin\psi}{\Lambda} \left(1+\frac{\dot{\psi}^2}{8\Xi^2}\right)\right] + \delta \dot{\psi} \left(\frac{\dot{\psi}}{4\Lambda\Xi^2} + \frac{\mu \dot{\psi}}{4\Lambda\Xi^2}  \cos{\psi} \right)\right\}\mathrm{d}\sigma}.}_{\displaystyle\phantom{.}\hspace{2.5cm}\left\llbracket\dfrac{\dot{\psi}}{4\Lambda\Xi^2}(1+\mu\cos{\psi})\,\delta\psi\right\rrbracket_{\sigma_\ast}^{1/2}-\int_{\sigma_\ast}^{1/2}{\left[\dfrac{\ddot{\psi}}{4\Lambda\Xi^2}(1+\mu\cos{\psi})+\dfrac{\mu\sin{\psi}}{\Lambda}\left(1-\dfrac{\dot{\psi}^2}{8\Xi^2}\right)\right]\delta\psi\,\mathrm{d}\sigma}}\hspace{0.65cm}\phantom{.}
\end{align}
\end{widetext}
The coefficient of the variation $\delta\sigma_\ast$ vanishes if $\dot{\psi}$ and $\psi$ are continuous across $\sigma=\sigma_\ast$. We therefore solve the governing equations that we read off the final term in the variation subject to the continuity conditions $\dot{\psi}(\sigma_\ast^+)=\Psi$ and $\psi(\sigma_\ast^+)=\Psi\sigma_\ast$, and subject to the condition $\dot{\psi}(1/2)=0$ discussed above. With these boundary conditions, the boundary variation above also vanishes.

For numerical purposes, it is convenient to write the equations in terms of the rescaled variable $\sigma'$ defined via
\begin{align}
\sigma=\sigma_\ast+(1-2\sigma_\ast)\sigma',
\end{align}
so that $\sigma_\ast\leq\sigma\leq 1/2\Longrightarrow 0\leq\sigma'\leq1/2$. With respect to $\sigma'$, the governing equation becomes
\begin{align}
\dv[2]{\psi}{\sigma'{}}+\frac{4\mu\Xi^2\sin{\psi}}{1+\mu\cos{\psi}}\left[(1-2\sigma_\ast)^2-\frac{1}{8\Xi^2}\left(\dv{\psi}{\sigma'}\right)^2\right]=0.   \label{eq:gov01}
\end{align}
The solution of this second-order differential equation and the three constants $\mu,\Lambda,\sigma_\ast$ are determined by the three boundary conditions
\begin{align}
&\psi(0)=\Psi\sigma_\ast,&&{\dv{\psi}{\sigma'{}}}(0)=(1-2\sigma_\ast)\Psi,&&{\dv{\psi}{\sigma'{}}}(1/2)=0,\label{eq:gov02}
\end{align}
and the two integral conditions
\begin{subequations}\label{eq:gov03}
\begin{align}
&\int_{0}^{1/2}{\dfrac{\cos{\psi}}{1-2\sigma_\ast}\left[(1-2\sigma_* )^2+\frac{1}{8\Xi^2} \left(\frac{\mathrm{d}\psi}{\mathrm{d}\sigma'}\right)^2\right]\mathrm{d}\sigma'}\nonumber\\
&\quad=\frac{\Lambda (1-D)}{2}- \frac{\sin(\Psi \sigma_* )}{\Psi}\left( 1+\frac{\Psi^2}{8\Xi^2}\right),\\
&\int_{0}^{1/2}{\frac{1}{8(1-2\sigma_\ast)\Lambda^2\Xi^2}\left(\dv{\psi}{\sigma'}\right)^2\mathrm{d}\sigma'}\nonumber\\
&\quad=\dfrac{1}{2}\left(1-\frac{1}{\Lambda^2}\right)+\left(\frac{\Psi}{4\Lambda\Xi^2}\pdv{\Psi}{\Lambda}-\frac{\Psi^2 }{8\Lambda^2\Xi^2}\right)\sigma_\ast\nonumber\\
&\qquad-\mu\left[ \frac{1-D}{2\Lambda}-\frac{\sigma_* \cos (\Psi \sigma_* )}{\Lambda \Psi} \pdv{\Psi}{\Lambda} \left(1+\frac{\Psi^2}{8\Xi^2}\right)\right.\nonumber\\
&\qquad\qquad\left.+\frac{ \sin (\sigma_*  \Psi)}{\Psi^2\Lambda}\pdv{\Psi}{\Lambda}\left( 1-\frac{\Psi^2}{8\Xi^2}\right)\right].
\end{align}
\end{subequations}
The solution of Eq.~\eqref{eq:gov01} subject to conditions~\eqref{eq:gov02} and \eqref{eq:gov03} determines the shape of the deformed monolayer~\cite{Note1}.
\section{Asymptotic analysis of ``unbuckling''}
In this section, we investigate the ``unbuckling'' bifurcation asymptotically by expanding the governing equations near the bifurcation in the limit of an epithelium of many cells, $\Xi\gg1$, or, equivalently, $\xi\equiv 1/\Xi\ll 1$. The bifurcation happens at a critical compression $D = D_\text{bif}$ and a critical value $\ell_0=\ell_{0,\text{bif}}$. Since $\Xi/\ell_0$ and $\ell_0$ are the nondimensional length and height of the undeformed monolayer, $r = \Xi / \ell_0^2$ is its undeformed aspect ratio. The relation $\smash{\ell_0^2=(r\xi)^{-1}}$ will allow us to eliminate $\ell_0$ and formulate the problem in terms of the critical value $r=r_\text{bif}$.

Now, as $\xi\to 0$, we expect that $D_\text{bif}\to D_0$ and $r_\text{bif}\to r_0$, for some $D_0,r_0$ to be determined. The structure of the equations suggests writing
\begin{align}
&D=D_0+\delta\xi^2,&&r=r_0+\rho\xi^2,\label{eq:Dr}
\end{align}
where $\delta,\rho=\mathcal{O}(1)$, so that the unbuckling bifurcation occurs at $\delta=\delta_\text{bif}$ and $\rho=\rho_\text{bif}$, to be determined in the asymptotic calculation. By varying $\delta$ and $\rho$, we probe different compressions and aspect ratios close to the unbuckling bifurcation; it turns out that these compressions are both sufficiently small to allow solving the governing equations in closed form and sufficiently large to explain the numerical observations for $D<D_\text{bif}$ as well as those for $D>D_\text{bif}$. We will in fact introduce later an inner expansion replacing Eq.~\eqref{eq:Dr} to match the different behaviours for $D<D_\text{bif}$ and $D>D_\text{bif}$.

Based on the numerical observations in the main text, we expect the unbuckling bifurcation to be associated with the state of the epithelium in which nearly all cells are triangular, i.e. $\sigma_\ast\approx 1/2$. We therefore introduce an expansion\begin{subequations}\label{eq:exp}
\begin{align}
\sigma_\ast  = \frac12 + s_1 \xi + s_2 \xi^2 + s_3\xi^3+ \mathcal{O}\bigl(\xi^4\bigr).\label{eq:sexp}
\end{align}
Similarly, we expand
\begin{align}
\Psi &= p_0 + p_1 \xi + p_2 \xi^2  + p_3\xi^3+\mathcal{O}\bigl(\xi^4\bigr),\\
\Lambda &= \Lambda_0 + \Lambda_1 \xi + \Lambda_2 \xi^2  + \Lambda_3\xi^3+\mathcal{O}\bigl(\xi^4\bigr), \label{eq:Lexp}\\
\mu &=  \mu_0 + \mu_1 \xi + \mu_2 \xi^2  + \mu_3\xi^3+\mathcal{O}\bigl(\xi^4\bigr), 
\end{align}
and
\begin{align}
\psi(\sigma') = \psi_0 + \psi_1(\sigma') \xi + \psi_2(\sigma') \xi^2  + \psi_3(\sigma')\xi^3+\mathcal{O}\bigl(\xi^4\bigr),
\end{align}
\end{subequations}
in which we assume that $\psi_0$ is a constant because we expect, using the first of Eqs.~\eqref{eq:gov02}, $\psi\approx \Psi\sigma_\ast$ for $\sigma_\ast\approx 1/2$. We shall insert these expansions into the governing equations and boundary conditions and solve these order-by-order, using \textsc{Mathematica} (Wolfram, Inc.) to manipulate complicated algebraic expressions. Substituting the definitions of $\xi$ and $r$ into Eqs.~\eqref{eq:gov04}, \eqref{eq:dpsiL}, \eqref{eq:gov01}--\eqref{eq:gov03}, we are thus to solve
\begin{widetext}
\begin{subequations}    
\begin{align}
\dv[2]{\psi}{\sigma'{}}=\frac{4\mu\sin{\psi}}{(1+\mu\cos{\psi})\xi^2}\left[\frac{\xi^2}{8}\left(\dv{\psi}{\sigma'}\right)^2-(1-2\sigma_\ast )^2\right], \label{gov1}
\end{align}
subject to
\begin{align}
&\psi(0) = \Psi \sigma_\ast, \label{gov2} \\
&\dv{\psi}{\sigma'}{(0)} = (1-2\sigma_\ast)\Psi, \label{gov3} \\
&\dv{\psi}{\sigma'}{(1/2)} = 0, \label{gov4}\\
&\int_{0}^{1/2}{\dfrac{\cos{\psi}}{1-2\sigma_\ast}\left[(1-2\sigma_\ast)^2+\frac{\xi^2}{8}\left(\frac{\mathrm{d}\psi}{\mathrm{d}\sigma'}\right)^2\right]\mathrm{d}\sigma'}=\frac{\Lambda (1-D)}{2}- \frac{\sin(\Psi  \sigma_* )}{\Psi }\left( 1+\frac{\Psi^2\xi^2}{8}\right), \label{gov5} \\
&\int_{0}^{1/2}{\frac{\xi^2}{8\Lambda^2(1-2\sigma_\ast)}\left(\dv{\psi}{\sigma'}\right)^2\mathrm{d}\sigma'} = \frac12\left(1-\frac{1}{\Lambda^2}\right)-\dfrac{3\sigma_\ast\Psi\xi^2}{8\Lambda^2}\dfrac{8\bigl(4r-3\Lambda^2\Psi\bigr)-5\Lambda^2\Psi^3\xi^2}{24\Lambda^2-\xi^2\Psi\bigl(4r+9\Lambda^2\Psi\bigr)}  \notag \\
&\qquad - \mu \left\{\frac{1-D}{2\Lambda}+\left[\frac{\sigma_* \cos (\Psi \sigma_* )}{\Lambda \Psi }\left(1+\frac{\Psi ^2\xi^2}{8}\right) -\frac{ \sin (\sigma_*  \Psi )}{\Psi ^2\Lambda}\left( 1-\frac{\Psi^2\xi^2}{8}\right)\right]\dfrac{24\bigl(2r+\Lambda^2\Psi\bigr)+\xi^2\Psi^2\bigl(2r-3\Lambda^2\Psi\bigr)}{\Lambda\bigl[24\Lambda^2-\xi^2\Psi\bigl(4r+9\Lambda^2\Psi\bigr)\bigr]}\right\},\label{gov6} \\
&\frac{1}{\Lambda}- \frac{\Lambda\Psi}{2r}+\frac{\xi^2\Psi^2}{24\Lambda}+\frac{\Lambda\xi^2\Psi^3}{16r}= 0. \label{gov7}
\end{align}
\end{subequations}
We will, in particular, want to obtain an expansion for another physically relevant observable, the buckling amplitude $A$ of the midpoint of the monolayer, from its vertical deflection $y(\sigma)$. By definition and by symmetry of the buckled shape, $A=y(1)=2y(1/2)$. The differential equation satisfied by $y(\sigma)$~\cite{Note1} was derived in Ref.~\cite{haas19}. Thus
\begin{align}
A&=2\int_0^{1/2}{\dfrac{\mathrm{d}y}{\mathrm{d}\sigma}\,\mathrm{d}\sigma}=\dfrac{2}{\Lambda}\sqrt{\dfrac{r}{\xi}}\int_0^{1/2}{\left[\left(1 + \frac{\dot{\psi}^2}{24 \Xi^2}\right)\sin{\psi}+\frac{\ddot{\psi}}{12 \Xi^2}\cos{\psi}\right]\mathrm{d}\sigma}\nonumber\\
&=\dfrac{2}{\Lambda}\sqrt{\dfrac{r}{\xi}}\left\{\frac{1-\cos(\Psi \sigma_* )}{\Psi}\left( 1+\xi^2 \frac{\Psi^2}{24}\right)+(1-2\sigma_\ast)\int_0^{1/2}{\left[\sin\psi +\frac{\xi^2}{12(1-2\sigma_\ast)^2} \left(\frac{\dot{\psi}^2}{2} \sin \psi + \ddot{\psi}\cos\psi \right) \right]\mathrm{d}\sigma'}\right\}.\label{eq:A}
\end{align}
\end{widetext}
Now, plugging the expansions~\eqref{eq:exp} into Eqs.~\eqref{gov2}, \eqref{gov5}, \eqref{gov6}, and \eqref{gov7}, we find, at leading order,
\begin{align}
&\Lambda_0=1,&&\psi_0 = r_0,&& p_0 = 2r_0,&& D_0 = 1-\frac{\sin{r_0}}{r_0}.\label{eq:lo}
\end{align}
These leading-order results are independent of $\delta$, and hence they apply both above and below $D_{\rm bif}$. The next-order terms will depend, however, on whether we are above or below the bifurcation, and so we will consider these two situations separately in what follows.

\subsection{Solution below the bifurcation: $\vec{D< D_\text{bif}}$}
We begin by assuming $s_1\neq 0$. This will turn out to correspond to $D< D_\text{bif}$. In this case, we immediately get $\mu_0 = 0$ from the governing equation \eqref{gov1} at lowest order. After substituting this and the leading order results~\eqref{eq:lo} into the governing equation~\eqref{gov1}, we obtain, at next order and using \textsc{Mathematica},
\begin{subequations}
\begin{equation}
     \frac{1}{4s_1^2} \dv[2]{\psi_1}{\sigma'{}} + 4\mu_1 \sin r_0 = 0,
\end{equation}
which can be integrated to yield
\begin{equation}
\psi_1(\sigma') = a_1 + \sigma' b_1 - 8s_1^2 \sigma'^2 \mu_1 \sin r_0,
\end{equation}
\end{subequations}
where $a_1,b_1$ are constants of integration. Equations~\eqref{gov3} and \eqref{gov4} then give
\begin{subequations}
\begin{align}
&b_1 + 4 r_0 s_1 = 0,&&b_1 - 8 s_1^2 \mu_1 \sin r_0 = 0,
\end{align}
which yields
\begin{align}
&b_1 = - 4 r_0 s_1,&&\mu_1 = - \frac{r_0}{2s_1 \sin r_0 }. \label{eq:b1mu1}
\end{align}
\end{subequations}
From Eqs.~\eqref{gov2} and \eqref{gov7}, we then obtain
\begin{align}
&p_1 = - 4 r_0 \Lambda_1,&&a_1 = 2r_0 (s_1 - \Lambda_1).
\end{align}
Equations \eqref{gov5} and \eqref{gov6} then reduce, at order $\mathcal{O}(\xi)$, to
\begin{subequations}
\begin{align}
\frac{\Lambda_1 (\sin r_0 - 2r_0 \cos r_0 )}{2r_0} &= 0,  \label{eq:tan_a}\\
\frac{2 r_0 \cot r_0 +4s_1 \Lambda_1 -1 }{2s_1} &= 0. \label{eq:tan_b}
\end{align}
\end{subequations}
Now, $\tan r_0 = 2 r_0 $  or $\Lambda_1 = 0$ from Eq.~\eqref{eq:tan_a}. If $\tan r_0 = 2 r_0 $, then Eq.~\eqref{eq:tan_b} implies $\Lambda_1 = 0$ since we assume that $s_1\neq0$. If $\Lambda_1 = 0$, then $\tan r_0 = 2 r_0$ from Eq.~\eqref{eq:tan_b}. Thus, in any case, we have
\begin{align}
&\Lambda_1 = 0,&&   \tan{r_0} = 2 r_0.\label{eq:tan}
\end{align}
This transcendental equation yields $r_0\approx 1.17$, whence, from Eqs.~\eqref{eq:lo}, ${D_0\approx 0.21}$. 

At this stage, $s_1$ remains undetermined. We therefore return to Eqs.~\eqref{gov6} and \eqref{gov7}, now at order $\mathcal{O}\bigl(\xi^2\bigr)$, to obtain
\begin{align}
&\Lambda_2 = \frac{9 r_0^2}{4},&&p_2 = \frac{6\rho - 23r_0^3}{3}.\label{eq:L2p2}
\end{align}
Boundary condition~\eqref{gov5} then gives
\begin{equation}
s_1^ 2 =  \frac{3\rho - r_0^3 - 3 r_0 \delta \sec{r_0}}{4r_0^3}.\label{eq:s1below}
\end{equation}
This expression yields a real value for $s_1$ if and only if ${\delta<\bigl(\rho-r_0^3/3\bigr)\cos{r_0}}$. The breakdown of this asymptotic solution for larger $\delta$ heralds the bifurcation. This indicates
\begin{align}
\delta_\text{bif} = \frac{\bigl(3\rho_\text{bif} - r_0 ^3 \bigr)\cos{r_0}}{3 r_0}.\label{eq:dbif0}
\end{align}
To determine $\delta_\text{bif}$ and $\rho_\text{bif}$, we will need another relation between these two quantities. This we will obtain below from the inner expansion announced earlier.

We conclude the asymptotic solution for $\delta<\delta_\text{bif}$ by calculating the corresponding buckling amplitude. We expand\begin{subequations}
\begin{align}
A=A_{-1/2}\xi^{-1/2}+A_{1/2}\xi^{1/2}+\mathcal{O}\bigl(\xi^{3/2}\bigr),
\end{align}
and, by substituting the above results into Eq.~\eqref{eq:A}, obtain
\begin{align}
A_{-1/2}&=\frac{1-\cos{r_0}}{\sqrt{r_0}},&&A_{1/2}=0.
\end{align}
\end{subequations}

\subsection{Solution above the bifurcation: $\vec{D> D_\text{bif}}$}
To find the asymptotic solution that will correspond to ${D> D_\text{bif}}$, we will assume instead that $s_1=0$. This is consistent with Eq.~\eqref{eq:s1below}, which shows that $s_1\to 0$ as $\delta\to\delta_\text{bif}^-$.

We will further assume that $s_2\not=0$; at the end of this section, we prove that $s_1=s_2=0$ leads to a contradiction. With this assumption, the governing equation~\eqref{gov1} at order $\mathcal{O}\bigl(\xi^{-3}\bigr)$ yields 
\begin{subequations}
\begin{equation}
-\frac{1}{4s_2^2} \dv[2]{\psi_1}{\sigma'{}} = 0\quad\text{subject to }\dv{\psi_1}{\sigma'}{}(0)=\dv{\psi_1}{\sigma'}{}(1/2)=0,
\end{equation}
using boundary conditions~\eqref{gov3} and \eqref{gov4} at order $\mathcal{O}(\xi)$. This shows that
\begin{equation}
\psi_1(\sigma') = a_1, 
\end{equation}
\end{subequations}
where $a_1$ is a constant of integration.  Equations \eqref{gov2} and \eqref{gov7} at order $\mathcal{O}(\xi)$ further yield
\begin{align}
& a_1 = - 2 r_0 \Lambda_1,&& p_1 = - 4 r_0 \Lambda_1.
\end{align}
With these results, Eq.~\eqref{gov5} gives
\begin{equation}
\frac{\Lambda_1}{2r_0}(\sin r_0-2 r_0 \cos r_0) =0 ,\label{eq:tan1}
\end{equation}
whereas Eq.~\eqref{gov6} leads to
\begin{equation}
\frac{\mu_0}{r_0} (\sin r_0 - 2r_0 \cos r_0 ) = 0. 
\label{eq:tan2}
\end{equation}
At this stage, we have not yet obtained $\mu_0$. We do note however that, since $s_1\to 0$ as $\delta\to\delta_\text{bif}^-$ and from the second of Eqs.~\eqref{eq:b1mu1}, $\mu_1\uparrow\infty$ as $\delta\to\delta_\text{bif}^-$. Hence we expect $\mu_0\neq 0$ here, for $\delta>\delta_\text{bif}$. We therefore need to take the asymptotic solution to higher order. At next order, the governing equation~\eqref{gov1} is
\begin{subequations}
\begin{equation}
\frac{1}{4s_2^2} \dv[2]{\psi_2}{\sigma'{}} + \frac{4 \mu_ 0 \sin r_0}{1+ \mu_0 \cos r_0} = 0,\label{eq:psi2eq}
\end{equation}
which, from Eqs.~\eqref{gov3} and \eqref{gov4}, is subject to
\begin{align}
&\dv{\psi_2}{\sigma'}{}(0)=-4 r_0 s_2,&& \dv{\psi_1}{\sigma'}{}(1/2)=0. \label{eq:psi2bc}
\end{align}
\end{subequations}
Now if $\mu_0  =  0$, then $\mathrm{d}\psi_2/\mathrm{d}\sigma' = \text{const.}$ from Eq.~\eqref{eq:psi2eq}, which is consistent with Eqs.~\eqref{eq:psi2bc} if and only if $s_2 = 0$, contradicting our assumption $s_2\neq 0$. Hence $\mu_0\neq 0$. From Eq.~\eqref{eq:tan2}, we can now conclude that 
\begin{equation}
\tan r_0 = 2r_0.\label{eq:tantan}
\end{equation}
This is, reassuringly, again the transcendental equation that we have already obtained for $\delta<\delta_\text{bif}$ as the second of Eqs.~\eqref{eq:tan}. With this, Eq.~\eqref{eq:tan1} is also now satisfied. 

Next, the solution of Eq.~\eqref{eq:psi2eq} subject to boundary conditions~\eqref{eq:psi2bc} is 
\begin{equation}
\psi_2(\sigma') = a_2 + b_2 \sigma' - \frac{8 s_2^2  \mu_0 \sin r_0 }{1+ \mu_0 \cos r_0}\sigma'^2,
\end{equation}
where $a_2,b_2$ are further constants of integration. Substituting this into boundary conditions \eqref{gov3} and \eqref{gov4} gives
\begin{align}
&b_2 = - 4 r_0 s_2, &&\mu_0 = - \frac{r_0}{r_0 \cos r_0 + 2 s_2 \sin r_0 }. \label{eq:b2mu0}
\end{align}
Plugging these expressions back into Eqs.~\eqref{gov2} and \eqref{gov7} leads to
\begin{subequations}
\begin{align}
a_2 &= \frac{2r_0^3}{3} + \rho + r_0 \bigl(2s_2 + 3 \Lambda_1^2 - 2\Lambda_2 \bigr),\\
p_2 &= \frac{4r_0^3}{3} + 2\rho + 6 r_0 \Lambda_1^2 - 4 r_0 \Lambda_2.
\end{align}
\end{subequations}
From Eq. \eqref{gov5} at order $\mathcal{O}\bigl(\xi^2\bigr)$, we then obtain
\begin{equation} 
\Lambda_1^2 = \frac{r_0^3 - 3 \rho + 3 r_0 \delta \sec r_0 }{3r_0\bigl(4r_0^2 - 1\bigr)}.\label{eq:L1}
\end{equation}
This expression results in a real value of $\Lambda_1$ if and only if ${\delta>\bigl(\rho-r_0^3/3\bigr)\cos{r_0}}$, and its breakdown for smaller $\delta$ again heralds the bifurcation. This shows that
\begin{equation}
\delta_\text{bif} = \frac{\bigl(3\rho_\text{bif} - r_0^3\bigr) \cos{r_0}}{3 r_0 }.\label{eq:dbif2}
\end{equation}
Reassuringly, we obtained the same value of $\delta_\text{bif}$ in Eq.~\eqref{eq:dbif0} from the expansion for $\delta<\delta_\text{bif}$. We also recall, from the first of Eqs.~\eqref{eq:tan}, that $\Lambda_1\equiv 0$ vanishes below the bifurcation ($\delta<\delta_\text{bif}$). This is consistent with Eq.~\eqref{eq:L1}, which implies that $\Lambda_1\to 0$ as $\delta\to\delta_\text{bif}^+$.

We still have not determined $\mu_0$ since its expression in Eqs.~\eqref{eq:b2mu0} depends on $s_2$. However, Eq.~\eqref{gov6} at order $\mathcal{O}(\xi)$ yields
\begin{subequations}
\begin{equation}
\frac{8\Lambda_1 \left(s_2 + r_0^2 \right)}{1+4s_2} = 0.
\end{equation}
Since we have just shown that $\Lambda_1\neq 0$, this implies
\begin{equation}
s_2 = - r_0^2.\label{eq:s2}
\end{equation}
\end{subequations}
With this, the second of Eqs.~\eqref{eq:b2mu0} reduces to
\begin{equation}
\mu_0 =  - \frac{\cos r_0}{\cos 2 r_0}.
\end{equation}
To determine the bifurcation behaviour of the compressive force, we need to determine $\mu_1$, for which purpose we need to take the expansion to still higher orders. Integrating the governing equation \eqref{gov1} at next order and simplifying using Eq.~\eqref{eq:tantan} gives
\begin{align}
\psi_3(\sigma') &= r_0\left[8\bigl(s_3+r_0^2\Lambda_1\bigr)-\mu_1\bigl(1-4r_0^2\bigr)^2\cos{r_0}\right] \sigma'^2\nonumber\\
&\quad+a_3+b_3\sigma',
\end{align}
where $a_3,b_3$ are more constants of integration. From boundary conditions \eqref{gov4} and \eqref{gov5}, we then obtain
\begin{align}
b_3&= - 4r_0\bigl(s_3 + 2r_0^2 \Lambda_1\bigr),&&\mu_1 = \frac{4s_3\sec{r_0}}{\bigl(4r_0^2-1\bigr)^2}.\label{eq:b3mu1}
\end{align}
Now, from Eq.~\eqref{gov6}, we find
\begin{align}
s_3=\dfrac{\left(72\Lambda_1^2\!+\!5\right)r_0^3-92r_0^5+12\bigl(2r_0^2\!-\!1\bigr)\rho+6r_0\bigl(\delta\sec{r_0}\!+\!\Lambda_1^2\bigr)}{48r_0\Lambda_1}.\label{eq:s3}
\end{align}
Substituting Eqs.~\eqref{eq:L1} and \eqref{eq:s3} into the second of Eqs.~\eqref{eq:b3mu1} finally yields
\begin{widetext}
\vspace{-10pt}
\begin{align}
\mu_1=\pm\dfrac{\bigl[136r_0^5-368r_0^7+3r_0^3\bigl(32\delta\sec{r_0}-1\bigr)+6\rho\bigl(1-24r_0^2+16r_0^4\bigr)\bigr]\sec{r_0}}{4\sqrt{3r_0}\bigl(4r_0^2-1\bigr)^{5/2}\bigl(r_0^3-3\rho+3r_0\delta\sec{r_0}\bigr)^{1/2}}.\label{eq:mu1}
\end{align}
\end{widetext}
This exhibits the two branches of the compressive force above the bifurcation, and completes the asymptotic solution for $\delta>\delta_\text{bif}$. Again, we still need to determine $\delta_\text{bif}$ and $\rho_\text{bif}$. These expressions, derived below, will be seen to regularise the apparent divergence of $\mu_1$ as $\delta\to\delta_\text{bif},\rho\to\rho_\text{bif}$.

We conclude this subsection by computing the buckling amplitude from Eq.~\eqref{eq:A}. Again, we expand
\begin{subequations}
\begin{align}
A=A_{-1/2}\xi^{-1/2}+A_{1/2}\xi^{1/2}+O\bigl(\xi^{3/2}\bigr),
\end{align}
to find
\begin{align}
A_{-1/2}&=\frac{1-\cos{r_0}}{\sqrt{r_0}},\\
A_{1/2}&=\mp\frac{\sec{r_0}-1}{r_0}\left(\dfrac{r_0^3-3\rho+3r_0\delta\sec{r_0}}{3\bigl(4r_0^2-1\bigr)}\right)^{1/2},\label{eq:A1}
\end{align}
\end{subequations}
which finally expresses the ``unbuckling'' instability analytically: On one of the two branches, the amplitude decreases with increasing compression.

\subsubsection*{\textbf{Inconsistency of expansions for $\vec{\sigma_\ast}$ with $\vec{s_1=s_2=0}$}}
In the above calculation for $\delta>\delta_\text{bif}$, we assumed that $s_1=0$ and $s_2\neq 0$, while, in the asymptotic calculation for $\delta<\delta_\text{bif}$, we assumed that $s_1\neq 0$. Here, we will prove that expansions that assume instead that $s_1=s_2=0$ lead to a contradiction.

We thus suppose that $\sigma_\ast=1/2+s_k\xi^k+\mathcal{O}\bigl(\xi^{k+1}\bigr)$, where $s_k\neq0$, for some integer $k>2$. The expansion of $\psi(\sigma')$ is then
\begin{align}
\psi(\sigma') = \psi_0 + \xi \psi_1(\sigma') + \cdots + \xi^k \psi_k(\sigma')+\mathcal{O}\bigl(\xi^{k+1}\bigr). 
\end{align}
We shall prove by strong induction that $\psi_0,\psi_1,\dots,\psi_\ell,\dots,\psi_k$ are constants. The base case $\ell=0$ holds by assumption. Suppose that $\psi_0,\psi_1,\dots,\psi_\ell$ are constants for $\ell<k$. Then the scalings of the different terms in the governing equation \eqref{gov1} are
\begin{align}
&\dv[2]{\psi}{\sigma'{}}=\frac{4\mu\sin{\psi}}{1+\mu\cos{\psi}}\left[\frac{1}{8}\left(\dv{\psi}{\sigma'}\right)^2-\dfrac{(1-2\sigma_\ast )^2}{\xi^2}\right],\\
&\phantom{.}\hspace{-3mm}\mathcal{O}\bigl(\xi^{\ell+1}\bigr)\hspace{3.5mm}\lesssim\mathcal{O}(1)\hspace{5.5mm}\mathcal{O}\bigl(\xi^{2\ell+2}\bigr)\hspace{6mm}\mathcal{O}\bigl(\xi^{2k-2}\bigr)\nonumber
\end{align}
Indeed, the first-term on the right-hand side is at most of order $\mathcal{O}(1)$ since $1+\mu\cos{\psi}>0$ because $\cos{\psi}\approx\cos{r_0}>0$ from the second of Eqs.~\eqref{eq:lo} and using $r_0\approx 1.17$ and $\mu>0$ in a physical solution in which the compressive force is positive. Since $\ell+1\leq k<2k-2$ as $k>2$ and $\ell<k$, it follows that
\begin{align}
\dv[2]{\psi_{\ell+1}}{\sigma'{}}=0 \implies \dv{\psi_{\ell+1}}{\sigma'{}}=\text{const.}.
\end{align}
Boundary condition~\eqref{gov4} shows that this constant vanishes, whence $\psi_{\ell+1}$ is constant, which completes the inductive step. Thus, by strong induction, $\psi_k$ is constant. However, boundary condition~\eqref{gov3} now yields
\begin{align}
\dv{\psi_k}{\sigma'{}}{}(0) = p_0 s_k,
\end{align}
From the leading order solution~\eqref{eq:lo}, $p_0 = 2 r_0\neq 0$, while $s_k\neq 0$ by assumption. This is a contradiction and proves our claim, that expansions with $s_1=s_2=0$ are inconsistent. We emphasise that we assumed, in the inductive step, that $k>2$ to neglect the last term in square brackets on the right-hand side of the governing equation. This explains why no such contradiction arose in the above calculation assuming $s_1=0,s_2\neq 0$.

\subsection{Inner expansion: Determining $\vec{\rho_\text{bif},\delta_\text{bif}}$}
Finally, to determine $\rho_\text{bif},\delta_\text{bif}$, we consider an inner asymptotic expansion that replaces Eqs.~\eqref{eq:Dr} with
\begin{align}
D=D_0+\delta_\text{bif}\xi^2+\Delta\xi^4,&&r=r_0+\rho_\text{bif}\xi^2+P\xi^4,
\end{align}
with $\Delta,P=\mathcal{O}(1)$ now controlling the distance from the bifurcation, and where, using Eqs.~\eqref{eq:lo}, \eqref{eq:tantan}, and \eqref{eq:dbif2},
\begin{align}
\tan{r_0}=2r_0,&&D_0=1-2\cos{r_0},&&\delta_\text{bif}=\dfrac{\bigl(3\rho_\text{bif}-r_0^3\bigr)\cos{r_0}}{3r_0}.
\end{align}
To match this inner expansion onto the asymptotic solutions for both $\delta>\delta_\text{bif}$ and $\delta<\delta_\text{bif}$, we replace Eqs.~\eqref{eq:sexp} and \eqref{eq:Lexp} with
\begin{subequations}
\begin{align}
\sigma_\ast&=\dfrac{1}{2}-r_0^2\xi^2+s_3\xi^3+\mathcal{O}\bigl(\xi^4\bigr),\\
\Lambda&=1+\dfrac{9r_0^2}{4}\xi^2+\Lambda_3\xi^3+\mathcal{O}\bigl(\xi^4\bigr),
\end{align}
\end{subequations}
where we have used the first of Eqs.~\eqref{eq:tan} and \eqref{eq:L2p2}, and Eq.~\eqref{eq:s2}. We leave the remaining expansion ansätze in Eqs.~\eqref{eq:exp} unchanged. 

The leading-order results~\eqref{eq:lo} then continue to hold, but Eq.~\eqref{gov7} at order $\mathcal{O}(\xi)$ now leads to $p_1 = 0$. The governing equation~\eqref{gov1} at lowest order is
\begin{equation}
\frac{1}{4r_0^4} \dv[2]{\psi_1}{\sigma'{}} = 0\quad\text{subject to }\dv{\psi_1}{\sigma'}{}(0)=\dv{\psi_1}{\sigma'}{}(1/2)=0,
\end{equation}
from boundary conditions \eqref{gov3} and \eqref{gov4}, and $\psi_1(0) = 0$, from boundary condition~\eqref{gov2}. This implies
\begin{equation}\psi_1(\sigma')=0.\end{equation}
Substitution into Eq.~\eqref{gov5} at order $\mathcal{O}\bigl(\xi^2\bigr)$ then leads to
\begin{equation}p_2 = - \frac{23}{3}r_0^3 + 2 \rho. \end{equation}
At next order, the governing equation \eqref{gov1} is 
\begin{subequations}
\begin{equation}\frac{1}{4 r_0^4}\dv[2]{\psi_2}{\sigma'{}}+\frac{4 \mu_0 \sin r_0}{1+\mu_0 \cos r_0 }  = 0. \end{equation}
This integrates to
\begin{equation} \psi_2(\sigma') = a_2 + b_2 \sigma' - \frac{8 r_0^4\mu_0 \sin r_0 }{1+ \mu_0 \cos r_0} \sigma'^2,\end{equation}
\end{subequations}
where $a_2,b_2$ are new constants of integration. Substituting this into Eqs.~\eqref{gov3} and \eqref{gov4}, we find
\begin{align}&b_2 = 4 r_0^3,&&\mu_0 = \frac{\sec{r_0}}{4r_0^2-1}.\end{align}
Equation~\eqref{gov6} at order $\mathcal{O}\bigl(\xi^2\bigr)$ finally yields
\begin{equation}
\rho_\text{bif} = \frac{r_0^3\bigl(92 r_0^2 - 3\bigr)}{6\bigl(4 r_0^2 - 1\bigr)}.\label{eq:rbif}\end{equation}
This determines $\rho_\text{bif}$ in terms of $r_0$ only, and also determines an expression for $\delta_\text{bif}$ depending only on $r_0$, viz., 
\begin{equation} 
\delta_\text{bif} = \frac{\bigl(84r_0^2-1\bigr)r_0^2\cos{r_0}}{6\bigl(4r_0^2 - 1\bigr)}.\label{eq:dbif}
\end{equation}
By substituting Eq.~\eqref{eq:rbif} into Eqs.~\eqref{eq:s1below}, we can now obtain an expression for $s_1$ at the critical aspect ratio $\rho=\rho_\text{bif}$,
\begin{align}
s_{1}=-\dfrac{\sqrt{3\sec{r_0}}}{2r_0}(\delta_\text{bif}-\delta)^{1/2}\quad\text{for $\delta<\delta_\text{bif}$},
\end{align}
using the requirement $s_1\leq 0$ for $\sigma_\ast\leq 1/2$. Similarly, from Eqs.~\eqref{eq:L1}, \eqref{eq:mu1}, \eqref{eq:A1}, we obtain expressions for $\Lambda_1$, $\mu_1$, $A_{1/2}$  at the critical aspect ratio $\rho=\rho_\text{bif}$,
\begin{subequations}
\begin{align}
\Lambda_1&=\pm\biggl(\dfrac{\sec{r_0}}{4r_0^2-1}\biggr)^{1/2}(\delta-\delta_\text{bif})^{1/2},\\
\mu_1&=\pm\dfrac{8r_0^2\sec^{3/2}{r_0}}{\bigl(4r_0^2-1\bigr)^{5/2}}(\delta-\delta_\text{bif})^{1/2},\\
A_{1/2}&=\mp\dfrac{\sqrt{\sec{r_0}}(\sec{r_0}-1)}{\sqrt{r_0}\bigl(4r_0^2-1\bigr)^{1/2}}(\delta-\delta_\text{bif})^{1/2}.
\end{align}
\end{subequations}

\begin{table*}[t]
\caption{Summary of asymptotic results. Asymptotic expansions and numerical approximations for different ``observable'' quantities, both below the bifurcation ($\delta<\delta_\text{bif}$) and above the bifurcation ($\delta>\delta_\text{bif}$).} \label{tab0}
\begin{ruledtabular}
\begin{tabular}{c}
{\begin{minipage}{17cm}\phantom{.}\vspace{-12pt}
\begin{align*}
&\text{observable}\hspace{25mm}\text{expansion for $\delta<\delta_\text{bif}$}\hspace{22mm}\text{expansion for $\delta>\delta_\text{bif}$\hspace{49mm}\phantom{.}}
\end{align*}\vspace{-14pt}\phantom{.}
\end{minipage}}\\
\hline
{\begin{minipage}{18cm}\phantom{.}\vspace{-6pt}
\begin{align*}
&\text{lateral side stretch}&\Lambda&=1+\mathcal{O}(\xi^2)&\Lambda&\sim1\pm\biggl(\dfrac{\sec{r_0}}{4r_0^2-1}\biggr)^{1/2}(\delta-\delta_\text{bif})^{1/2}\xi\\
&&&&&\approx 1\pm 0.756(D-D_\text{bif})^{1/2}\\[2mm]
&\text{compressive force}&\mu&=\mathcal{O}(\xi)&\mu&\sim-\frac{\cos r_0}{\cos 2 r_0}\pm\dfrac{8r_0^2\sec^{3/2}{r_0}}{\bigl(4r_0^2-1\bigr)^{5/2}}(\delta-\delta_\text{bif})^{1/2}\xi\\
&&&&&\approx0.572\pm 1.060(D-D_\text{bif})^{1/2}\\[2mm]
&\text{amplitude}&A&\sim\dfrac{1-\cos{r_0}}{\sqrt{r_0}}\xi^{-1/2}&A&\sim\dfrac{1-\cos{r_0}}{\sqrt{r_0}}\xi^{-1/2}\mp\dfrac{\sqrt{\sec{r_0}}(\sec{r_0}-1)}{\sqrt{r_0}\bigl(4r_0^2-1\bigr)^{1/2}}(\delta-\delta_\text{bif})^{1/2}\xi^{1/2}\\
&&&\approx 0.561 \xi^{-1/2}&&\approx\xi^{-1/2} \left[ 0.561 \mp 1.076(D-D_\text{bif})^{1/2}\right]\\[2mm]
&\text{extent of triangular cells}&\sigma_\ast&\sim\dfrac{1}{2} -\frac{\sqrt{3 \sec{r_0}}}{2r_0}(\delta_\text{bif}-\delta)^{1/2}\xi&\sigma_\ast&=\dfrac{1}{2}+\mathcal{O}\bigl(\xi^2\bigr)\\
&&&\approx \dfrac{1}{2} -1.183 (D_\text{bif}-D)^{1/2}&&
\end{align*}
\vspace{-8pt}\phantom{.}
\end{minipage}}
\end{tabular}
\end{ruledtabular}
\end{table*}

\begin{figure*}
\centering 
\includegraphics[width=17.8cm]{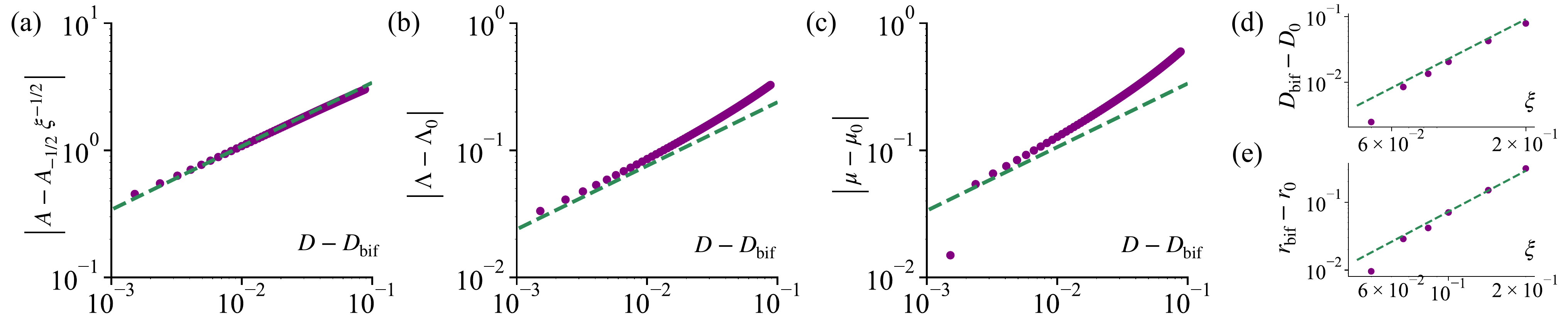}
\caption{Further comparison of numerical results (dot markers) and asymptotic results (dashed lines) near the bifurcation. (a)~Buckling amplitude: plot, for the ``buckling'' branch, of the absolute difference of the amplitude $A$ and its leading-order expansion $\smash{A_{1/2}\xi^{-1/2}}$ against $D-D_{\text{bif}}$ for $D>D_\text{bif}$. (b) Stretch of the lateral sides: plot of $|\Lambda-\Lambda_0|$ on the ``buckling'' branch against $D-D_{\text{bif}}$ for $D>D_\text{bif}$. (c) Compressive force: plot of $|\mu-\mu_0|$ on the ``buckling'' branch against $D-D_\text{bif}$ for $D>D_\text{bif}$. (d) Critical compression for the bifurcation: plot of $D_\text{bif}-D_0$ against $\xi$. (d) Critical aspect ratio for the bifurcation: plot of $r_\text{bif}-r_0$ against $\xi$.}
\label{fig:compare}
\end{figure*}

\subsection{Comparison with numerics}
We begin this subsection by summarising our results from the asymptotic calculations in Table \ref{tab0}. The value of $r_0$ determining the prefactors satisfies
\begin{equation} 
\tan{r_0} = 2r_0 \implies r_0 \approx 1.165.
\end{equation}
Using Eqs.~\eqref{eq:lo}, \eqref{eq:rbif}, 
\eqref{eq:dbif},  we find the compression and the aspect ratio at which the bifurcation occurs,   
\begin{subequations}\label{eq:bifpos}
\begin{align} 
D_\text{bif}&\sim 1\!-\!\dfrac{\sin{r_0}}{r_0}+\xi^2\frac{\bigl(84r_0^2\!-\!1\bigr)r_0^2\cos{r_0}}{6\bigl(4r_0^2 - 1\bigr)} \approx 0.212+2.277\xi^2,\\
r_\text{bif} &\sim r_0+\xi^2\frac{r_0^3\bigl(92 r_0^2 - 3\bigr)}{6\bigl(4 r_0^2 - 1\bigr)} \approx 1.165+7.260\xi^2.
\end{align}
\end{subequations}
Figure 3(d) of the main text compares the asymptotic results for $f=2\sigma_\ast$ from Table~\ref{tab0} with the numerical results from the continuum model near the bifurcation. The asymptotic scalings for the remaining observables $A,\Lambda,\mu$ are verified in \textfigref{fig:compare}{a)--(c}, which confirms that the numerical results approach the asymptotic ones as $D\to D_\text{bif}$. In \textfigref{fig:compare}{d} and \textfigref{fig:compare}{e}, we also verify the expressions~\eqref{eq:bifpos} for the position, in parameter space, of the bifurcation; the relation between $D_\text{bif}-D_0$ and $r_\text{bif}-r_0$ is also verified in Fig.~3(c) of the main text. As $D\to D_\text{bif}$ or $\xi\to 0$, numerical solution of the continuum equations becomes more difficult because the governing are singular at $\sigma_\ast=1/2$. This explains why the numerical results closest to $D=D_\text{bif}$ or $\xi=0$ agree less well with the asymptotic approximation.

\bibliography{main}